# Effects of Environment on the Size Evolution of Quiescent Galaxies: Comparing Galaxies in Clusters and in the Field at Two Rest-frame Wavelengths

Angelo George[1], Ivana Damjanov[1,7], Marcin Sawicki[1], Devin J. Williams[1], Lingjian Chen[1], Guillaume Desprez[1], Marianna Annunziatella[2], Stéphane Arnouts[3], Stephen Gwyn[4], Danilo Marchesini[5], Thibaud Moutard[6], and Anna Sajina[5]

[1] Institute for Computational Astrophysics and Department of Astronomy & Physics, Saint Mary's University, 923 Robie Street, Halifax, Nova Scotia, B3H 3C3, Canada; angelogeorget@gmail.com
[2] Centro de Astrobiología (CSIC-INTA), Ctra de Torrejón a Ajalvir, km 4, E-28850 Torrejón de Ardoz, Madrid, Spain
[3] Aix-Marseille University, CNRS, CNES, LAM, Marseille, France
[4] NRC Herzberg Astronomy and Astrophysics, 5071 West Saanich Road, Victoria, British Columbia, V9E 2E7, Canada
[5] Department of Physics & Astronomy, Tufts University School of Arts and Sciences, 574 Boston Avenue, Medford, MA 02155, USA
[6] European Space Agency (ESA), European Space Astronomy Centre (ESAC), Camino Bajo del Castillo s/n, 28692 Villanueva de laCañada, Madrid, Spain

## Abstract

We investigate the impact of environment on quiescent galaxy (QG) size evolution using the CLAUDS+HSC imaging covering 18.6 $\deg^2$ in five broad filters (*Ugriz*) and the effective radius of a single-Sérsic fit as a proxy for galaxy size. We estimate sizes in two rest-frame wavelengths—3000 Å (UV) and 5000 Å (optical)—for ~86,000 massive ($M_* \leqslant 10^{9.5} M_\odot$) field QGs and for 1000 of their similarly massive counterparts from 47 clusters at $0.1 < z < 0.85$. We fit the size–mass relation (SMR) for field and cluster QGs in five $\Delta z = 0.15$ redshift bins and use the characteristic size of $M_* = 5 \times 10^{10} M_\odot$ QGs (SMR's zero-point) to trace the change in galaxy size over cosmic time and in two types of environment. Sizes of QGs are larger in the rest-frame UV than in the rest-frame optical in both clusters and the field, and this difference is more prominent in the field sample. However, QGs in clusters are systematically smaller than the field QGs, and this difference is significantly more pronounced if measured in the rest-frame UV light. Modeling of the redshift evolution in the characteristic QG size as $R_e \propto (1+z)^\beta$ shows that the cluster QGs ($\beta = -1.02$ in UV and $\beta = -1.00$ in optical) grow in size as fast as the field QGs ($\beta = -0.95$ in UV and $-1.22$ in optical). This fast growth of cluster QGs is consistent with size increase driven by the accretion of two subpopulations onto clusters: (a) field QGs that are larger than their quiescent counterparts in clusters, and (b) environmentally quenched galaxies (newcomers) that are larger than the rest of the quiescent population.

## 1. Introduction

The structural properties of galaxies encode critical information about their formation and evolutionary histories. The size–mass relation (SMR), linking galaxy physical sizes to their stellar mass, has emerged as one of the fundamental scaling relations that provides insights into the interplay between stellar mass and structural growth over cosmic time. The SMR for quiescent galaxies (QGs) and star-forming galaxies (SFGs) differs markedly, with QGs generally being smaller than their SFG counterparts at fixed stellar mass. This dichotomy encodes the difference in evolutionary pathways between the two populations. The processes that drive mass assembly and quenching of star formation and produce the distinct stellar mass–size relations for the two populations include feedback, quenching, and mergers (e.g., S. Shen et al. 2003; I. Trujillo et al. 2004; A. van der Wel et al. 2014; L. A. Mowla et al. 2019; L. Kawinwanichakij et al. 2021; A. George et al. 2024).

Over the last two decades, the extensively studied evolution of the SMR has revealed significant size growth for galaxies at fixed stellar mass over the broadest redshift range, $0 < z < 5.5$ (e.g., I. Trujillo et al. 2004; R. J. Williams et al. 2010; A. van der Wel et al. 2014; N. Roy et al. 2018; K. A. Suess et al. 2019; L. Kawinwanichakij et al. 2021; L. Yang et al. 2021; A. George et al. 2024; M. Martorano et al. 2024; E. Ward et al. 2024). For QGs, this size evolution is primarily attributed to dry minor mergers, which deposit stars on galaxy outskirts, and internal dynamical processes that redistribute stellar mass like adiabatic expansion (e.g., L. Fan et al. 2008; I. Damjanov et al. 2009, 2023; A. B. Newman et al. 2012; L. Oser et al. 2012; A. Beifiori et al. 2014; J. R. Ownsworth et al. 2014; S. Belli et al. 2015; F. Buitrago et al. 2017; H. J. Zahid et al. 2019; M. L. Hamadouche et al. 2022). In addition, the average size evolution for the QG population is also affected by the newly quenched larger galaxies that are being added to the quiescent population (progenitor bias; e.g., P. G. van Dokkum & M. Franx 2001; R. P. Saglia et al. 2010; C. M. Carollo et al. 2013; S. Genel et al. 2018; I. Damjanov et al. 2023; A. George et al. 2024). Although these processes are well characterized in field environments, the role of dense regions (e.g., galaxy cluster cores) in modulating the SMR remains less clear.

Galaxy clusters represent some of the most extreme environments in the Universe, hosting thousands of galaxies within a single gravitational potential well and permeated by a hot, dense intracluster medium. Due to the high relative velocities of cluster galaxies, galaxy mergers are rare in cluster cores (D. Merritt 1985; A. G. Delahaye et al. 2017). As a result, the growth of individual QGs in clusters may not be as efficient as it is in the field (I. Damjanov et al. 2015).

[7] Canada Research Chair.

Nevertheless, galaxies in clusters are subject to a wide range of environmental processes, including ram pressure stripping, tidal forces, and violent galaxy–galaxy interactions. These processes alter galaxy star formation activity, dynamics, and (consequently) morphology (e.g., J. E. Gunn & J. R. Gott 1972; A. Dressler 1980; R. B. Larson et al. 1980; B. Moore et al. 1996, 1998; R. Feldmann et al. 2010; M. Hirschmann et al. 2014; Y. L. Jaffé et al. 2015; G. Castignani et al. 2020; N. Chamba et al. 2024). Such processes are particularly pronounced in cluster cores, where high galaxy densities and strong tidal forces dominate, potentially leading to morphological transformations not typically observed in the field. For example, tidal stripping can remove materials from galaxies, resulting in mass loss and structural changes (e.g., O. Y. Gnedin 2003; C. Blom et al. 2014; E. L. Łokas 2020).

Although a growing body of literature suggests that the environment affects galaxy evolution (e.g., Y.-j. Peng et al. 2010; Y. Ding et al. 2024), their imprints on the QG SMR are still contested. While some studies indicate that the cluster and field QGs have similar sizes (A. Rettura et al. 2010; A. B. Newman et al. 2014; T. Morishita et al. 2017; S. M. Sweet et al. 2017), many other works suggest the opposite. Some studies found that cluster QGs are larger than field QGs (M. C. Cooper et al. 2012; C. Papovich et al. 2012; R. Bassett et al. 2013; C. Lani et al. 2013; L. Delaye et al. 2014; S. Andreon 2018; J. C. C. Chan et al. 2018; A. V. Afanasiev et al. 2023), while some others find that QGs are on average smaller in clusters than in the field (T. Valentinuzzi et al. 2010a, 2010b; B. M. Poggianti et al. 2012; A. Raichoor et al. 2012; M. Cebrián & I. Trujillo 2014; J. Matharu et al. 2019). This discrepancy could arise from a variety of reasons, such as differences in the selection of cluster members, explored regions within clusters, mass and redshift ranges, wavelengths probed, and methodology for size measurements between studies.

Therefore, there exist significant gaps in our understanding of how cluster-specific processes influence galaxy structure and how this varies across different redshifts and stellar masses. Moreover, the scatter in the SMR may be influenced by the diversity of quenching pathways in clusters, from fast-acting mechanisms such as ram pressure stripping (T. Brown et al. 2017; A. Singh et al. 2019) to more gradual processes like starvation or strangulation (R. B. Larson et al. 1980; J. Trussler et al. 2021).

The aspect of quiescent galaxy size growth that has not been probed so far as a function of the environment is the change in growth rate with the observed rest-frame wavelength. Extensive studies have established that galaxy light profiles are generally more extended in shorter wavelengths than in longer ones (e.g., N. Tamura & K. Ohta 2000; F. La Barbera et al. 2005; A. van der Wel et al. 2014; R. Lange et al. 2015; V. Marian et al. 2018; K. A. Suess et al. 2019; L. Kawinwanichakij et al. 2021; A. George et al. 2024). This negative color gradient indicates bluer outskirts and redder inner regions of quenched galaxies.

A major reason for this blue-red stellar color dichotomy in galaxies is the age difference. For example, the UV region of a galaxy spectrum is dominated by the light emitted by young massive stars, whereas the light from the old low-mass stars dominates the optical/near-IR region (G. Bruzual & S. Charlot 2003). Additionally, bluer colors can also be a result of low-metallicity stars (age–metallicity degeneracy; G. Worthey 1994; B. Tang & G. Worthey 2013; P. de Meulenaer et al. 2014). Hence, the bluer colors signal more recently formed stars (or low-metallicity stars) than the redder colors.

A suite of studies show that QGs in the field grow through minor mergers and accretion (e.g., R. Bezanson et al. 2009; H. J. Zahid et al. 2019; J. Matharu et al. 2019, 2020; A. George et al. 2024). Because newly accreted stars can be young and/or have low metallicity, it should result in a negative color gradient (K. A. Suess et al. 2023). However, due to the high peculiar velocities of cluster galaxies, rich cluster cores are not conducive to merger-driven growth in QGs (D. Merritt 1985; A. G. Delahaye et al. 2017). Therefore, devoid of mergers, cluster QGs should have shallower color gradients and smaller UV sizes than field QGs.

In this work, we aim to bridge these gaps by examining the SMR of quiescent galaxies in field and extreme cluster environments (i.e., cluster cores) across redshift and stellar mass. We utilize deep imaging of 18.6 $\mathrm{deg}^2$ of the sky from the Canada–France–Hawaii Telescope's (CFHT's) CLAUDS (M. Sawicki et al. 2019) and the Subaru Telescope's Hyper Suprime-Cam Subaru Strategic Program (HSC-SSP; H. Aihara et al. 2018a, 2018b, 2019, 2022). This survey provides (1) a deep data set (median 5$\sigma$ depth of $U = 27.09$ and $i = 26.9$) to model the light profiles of galaxies; (2) a large sample from a wide area in the sky to select a significant number of clusters and to reduce the effect of cosmic variance.

Using a photometric data set that includes both rest-frame UV (3000 Å) and optical (5000 Å) wavelengths for galaxies at $0.1 < z < 0.85$, we investigate how environmental effects shape the size evolution of galaxies. This work follows the methodology we previously implemented in our pilot study (A. George et al. 2024, hereafter G+24) in which we explored the size evolution of SFGs and QGs from the central COSMOS field (1.6 $\mathrm{deg}^2$) in the two rest-frame wavelengths. Here we explore whether and how the size–mass relation slope, scatter, and wavelength dependence for QGs in cluster cores differs from the equivalent scaling relation for their field counterparts. By focusing both on the rest-frame UV and optical wavelengths, we also address the variation in the size of quiescent cluster members with stellar population properties (age and/or metallicity) and compare it with the results in the field. Our analysis thus traces the drivers of the average growth in the size of QGs that reside in the densest regions of the Universe, differentiating between the effects of quenching at the population level and processes that transform individual QGs after quenching.

The plan of this paper is as follows. In Section 2 we provide an overview of the CLAUDS+HSC survey, the data and auxiliary data products that we use, sample selection, classification of galaxies into SFGs and QGs, and separation of QGs into field and cluster samples. In Section 3.1, we present details of the methodology for 2D galaxy light profile fitting, simulations we perform to estimate systematic uncertainties in our size measurements, and the estimation of the Sérsic parameters in two rest-frame wavelengths. In Section 4, we describe the analytic fits to the SMR and present the results of the SMR fitting and size evolution for field and cluster QGs in two rest-frame wavelengths. We discuss the implications of our findings in Section 5. We summarize the main conclusions based on our analysis in Section 6.

Throughout this study, we use $M$ to denote stellar mass in solar units (i.e., $M = M_\star/M_\odot$) unless otherwise stated. We also use the AB magnitude system and adopt a standard cosmology with $\Omega_M = 0.3$, $\Omega_\Lambda = 0.7$, and $h = 0.7$ throughout this paper. Additionally, our mass measurements assume the Chabrier initial mass function (IMF; G. Chabrier 2003).

## 2. Data

### 2.1. Imaging, Photometry, and Galaxy Masses

We use the Deep and UltraDeep multiband ($U$ + griz) imaging data from CFHT Large Area $U$-band Deep Survey[8] (CLAUDS; M. Sawicki et al. 2019) and the HSC-SSP[9] (H. Aihara et al. 2018a, 2022). In our earlier work (G+24), we focused on the central 1.6 deg$^2$ of the COSMOS field because it is a well-studied region in the sky with rich ancillary data. By extending our analysis to the entire 18.6 deg$^2$ of the CLAUDS+HSC survey in four different fields (E-COSMOS, DEEP2-3, ELAIS-N1, and XMM-LSS), we can study rare environments such as cores of rich clusters.

The survey has a median 5$\sigma$ depth of 26.9 mag in the $i$ band for point sources, and this depth increases to 27.8 mag in the UltraDeep regions of the survey. Other bands also have comparable depths (see M. Sawicki et al. 2019; H. Aihara et al. 2022 for details). As was shown in G+24, the depth of the CLAUDS+HSC imaging is sufficient to analyze the morphology of massive galaxies even to their dim outskirts. The CLAUDS+HSC *Ugrizy* imaging is complemented by the SHIRAZ Survey (M. Annunziatella et al. 2023; IRAC 3.6 $\mu$m and 4.5 $\mu$m) in E-COSMOS, DEEP2-3, and ELAIS-N1 fields, and VIRCAM near-IR wavelength coverage ($Y$, $J$, $H$, and $K_s$ bands) from the VIDEO (M. J. Jarvis et al. 2013) and UltraVISTA (H. J. McCracken et al. 2012) over a combined 5.5 deg$^2$ across the E-COSMOS and XMM-LSS fields.

We select our galaxies from the CLAUDS+HSC photometric catalog by G. Desprez et al. (2023) made with the HSC Pipeline (hscPipe; J. Bosch et al. 2018). They estimate the photometric redshifts using Phosphoros[10] (Euclid Collaboration et al. 2020; S. Paltani et al. 2025, in preparation), a template-fitting code created to be part of the Euclid photo-$z$ pipeline. The catalog utilizes *Ugrizy* or *Ugrizy*+$JHK_s$ (where VISTA $JHK_s$ bands are available). The photo-$z$ measurements at $i < 25$ have a precision of 0.05 and an outlier fraction <6% (G. Desprez et al. 2023).

Photometric spectral energy distribution (SED) fitting is used to estimate the physical properties of galaxies such as stellar masses, star formation rates (SFRs), and rest-frame colors. The details will be presented in L. Chen et al. (2025, in preparation); briefly, these properties are estimated by applying the LePhare (O. Ilbert et al. 2006) template-fitting code to the *Ugrizy* photometry together with VISTA $JHK_s$ (in the XMM-LSS field) or IRAC Ch1 and Ch2 (in the other three fields).

The SED fitting is performed by fixing the redshifts of galaxies to the photometric redshift values from G. Desprez et al. (2023) and assuming the Chabrier IMF (G. Chabrier 2003) and an exponentially decaying star formation history. Comparing our stellar mass measurements to those from the COSMOS2020 Farmer+LePhare catalog (J. R. Weaver et al. 2022), we find that, at $0.1 < z < 0.85$, the median offset in log stellar mass between the measurements ranges from 0.01 dex at the low-mass end to 0.12 at $\log M \sim 11.5$, and the scatter is about 0.15–0.2 dex at all masses. This catalog by L. Chen et al. (2025, in preparation) contains more than 5 million $i < 26$ objects from the CLAUDS+HSC survey.

### 2.2. Selection of Quiescent Galaxies

We use a modified variant of the *NUVrK* color–color selection (S. Arnouts et al. 2013; T. Moutard et al. 2018, 2020) to identify QGs in our sample (L. Chen et al. 2025, in preparation). For this, we extend the standard *NUVrK* color–color selection method to a 4D space of color–color-redshift-mass to smoothly account for redshift and mass dependence. This redshift- and mass-dependent *NUVrK* QG selection is parameterized as

$$NUV - r > (c_1 z + c_2 \log M + c_3)[(r - K) + c_4],$$
$$NUV - r > c_5. \quad (1)$$

The values of the constants $c_1 - c_5$ are determined empirically to be 3.868, 2.397, 1.003, −0.278, and 0.039, respectively, by calibrating Equation (1) with reference to COSMOS2020 quiescent galaxies from the J. R. Weaver et al. (2023) catalog. For calibration, we took COSMOS2020 galaxies with logsSFR $< -11$ as QGs, and optimized the *NUVrK* selection boundary (Equation (1)) to reduce contamination by SF galaxies while increasing QG completeness. Full details of the SFG/QG classification method for the CLAUDS+HSC sample will be available in L. Chen et al. (2025, in preparation).

### 2.3. Cluster Selection

Several galaxy cluster/group catalogs are available for the survey region in the literature (e.g., M. Oguri et al. 2018; Z. L. Wen & J. L. Han 2021). These catalogs rely on photometric redshifts to identify cluster members. However, the photometric redshift estimates differ between these catalogs and our work because of the differences in methodology and the number/distribution of photometric bands used to derive photo-$z$'s. Unlike in these catalogs, the catalog G. Desprez et al. (2023) used to derive photo-$z$'s for this and other CLAUDS+HSC-based studies (e.g., D. J. Williams et al. 2024) include the CFHT $U$ band during photo-$z$ fitting, which significantly improves photometric redshift performance out to $z \sim 0.75$ (M. Sawicki et al. 2019). We therefore do not rely on the membership from available catalogs and select our own cluster galaxy sample.

We begin with the M. Oguri et al. (2018) catalog. Although they select the clusters using photo-$z$'s, such catalogs of rich clusters show good agreement with X-ray cluster catalogs (M. Oguri et al. 2018; N. Krefting et al. 2020). Because this catalog contains spectroscopic redshifts of the brightest cluster galaxies (BCGs), we assume the positions of these BCGs as the centers of clusters in our data.

We then search our photo-$z$ catalog for QGs in the vicinity of these BCGs; we pre-select those that lie within a cylindrical volume with a comoving radius of 3 Mpc at the redshift of the BCG and a length of $2\Delta z$ (where $\Delta z = 0.05(1 + z)$) centered on the BCG. The length of the cylinder reflects the redshift-dependent scatter in our photometric redshifts (see G. Desprez et al. 2023). These pre-selected QGs constitute potential cluster members within a cylinder centered on the spectroscopically confirmed BCG and with the cross-sectional radius of 3 cMpc.

[8] https://www.clauds.net/

[9] https://hsc.mtk.nao.ac.jp/ssp/

[10] https://anaconda.org/astrorama/phosphoros

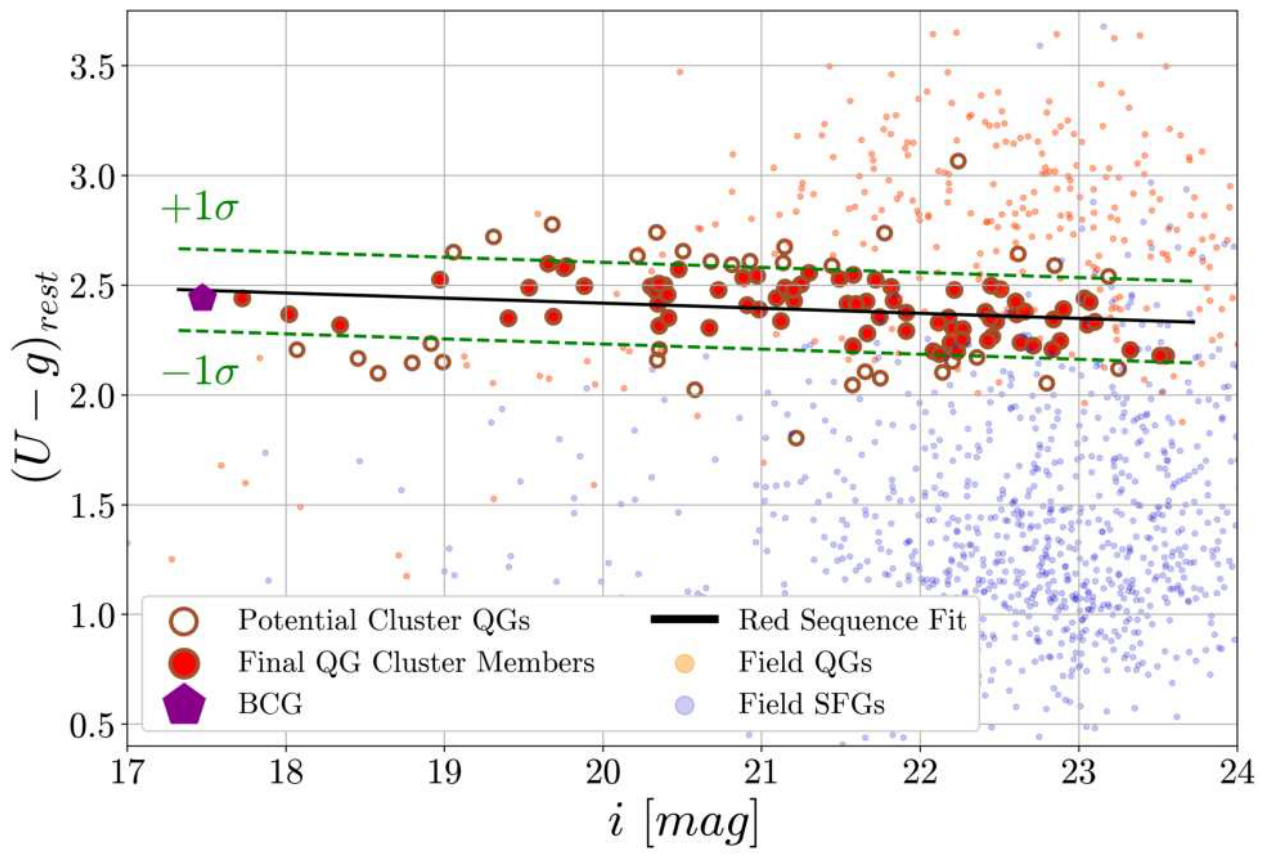

**Figure 1.** An example of red sequence selection of cluster members. The potential member QGs of a cluster in E-COSMOS ($z = 0.5479$) are shown as brown circles and their best-fit red sequence as a black line. The $1\sigma$ scatter around this red sequence is shown as green dashed lines. For purity, we restrict our cluster member selection to this $1\sigma$ region (filled red circles). The faint orange and blue dots in the figure denote foreground/background QGs and SFGs, respectively, within a cluster-centric radius, $r_c < 3$ cMpc from the BCG (purple pentagon), but outside redshift range (i.e., outside cylinder length, $2\Delta z$).

We then refine our selection of cluster members using the red sequence method (e.g., M. D. Gladders & H. K. C. Yee 2000; B. P. Koester et al. 2007; T. Valentinuzzi et al. 2011; E. S. Rykoff et al. 2014; M. Vakili et al. 2023) akin to the methodology implemented by M. Oguri (2014). Within the potential cluster members, we identify galaxies that lie on the sequence with almost identical red (rest frame at cluster redshift) color in the $U$-$g$ versus $i$ color–magnitude diagram (cluster red sequence).

Figure 1 shows the refinement of members of a sample cluster at $z = 0.5479$ using the color–magnitude diagram. The orange (blue) data points show the 320 QGs (969 SFGs) within 3 cMpc from the cluster BCG. We identify 116 QGs (brown circles) as potential cluster members through redshift selection ($2\Delta z$). After applying the red sequence refinement, we confirm the cluster membership of 81 QGs (filled red circles).

The cluster member finding algorithm we employ here is an iterative procedure. We draw a set of 100 photometric redshifts for each individual CLAUDS+HSC galaxy based on their redshift probability distribution functions (PDFs). We then run the cylindrical volume search and the red sequence selection for all draws independently. This approach enables us to calculate the cluster membership probability for every member galaxy based on the fraction of iterations in which they are confirmed to be cluster members. We adopt a membership probability >0.5 when selecting the final set of cluster members (i.e., a galaxy is a cluster member if it is confirmed, based on its photometric redshift draws, in more than 50% of the iterations).

Although our selection of cluster members extends to 3 cMpc from the BCG in the projected space, most of our analysis in this work is limited to inner regions of clusters (within the projected 1 cMpc) to explore the core regions of clusters where the density is the highest. Unless otherwise stated, we refer to galaxies within this projected inner region as cluster members. Our sample includes 47 clusters, 1035 cluster member QGs ($r_c < 1$ Mpc), and 2142 QGs in the cluster outskirts ($r_c$ : 1–3 Mpc). Figure 2 shows the sky and redshift positions of cluster QGs identified by our method in the CLAUDS+HSC fields. We classify all other QGs as field QGs. These also include all QGs that reside between projected 1 and 3 Mpc from the cluster centers (cluster outskirt QGs, fainter data points in Figure 2).

### 2.4. Final Field and Cluster Sample Selection

Motivated by the results of the simulations we perform (see G+24, and Section 3.2), we analyze the morphologies of galaxies brighter than $i = 24$. When fitting in the other bands—$U$, $g$, $r$, and $z$—we adopt additional magnitude cuts at 25.5, 25, 24.5, and 24 mag, respectively. Because we wish to have size measurements on both sides of the 4000 Å break, we limit our analysis to the redshift range of $0.1 < z < 0.85$, where the break is bracketed by our *Ugrizy* images.

We further restrict our analysis to massive galaxies ($\log M > 9.5$) for mass completeness of the sample. Figure 3 shows the stellar mass versus redshift diagram for all QGs (red) from the parent catalog. We show the 90% completeness limit as a white curve, estimated using the approach of L. Pozzetti et al. (2010), after introducing the magnitude cut of $i < 24$ (see Section 3.2 for the magnitude cut). At $\log M > 9.5$, our sample is 90% complete at $z \leqslant 0.55$. We use this subsample to constrain the pivot point of the SMR for QGs in our fields ($\log = 10.4$; see Section 4 and Appendix B for details of the pivot point estimation). At $z \leqslant 0.85$, our morphological sample is mass-complete above this pivot point.

Hence, we have selected 42,011 QGs with $\log M > 9.5$ over the redshift range $0.1 < z < 0.55$ and 30,246 QGs with $\log M > 10.4$ over the redshift range $0.55 < z < 0.85$. We identify 645 QGs from the former group and 229 QGs from the latter as cluster members residing in the cores ($r_c < 1$ cMpc) as discussed in Section 2.3. We consider the remaining as field QGs. At $\log M > 10.4$, we have 54,152 field QGs and 575 cluster QGs. Table 1 summarizes our selection cuts.

## 3. Methodology

### 3.1. Profile Fitting

We model the light profiles of the galaxies using a single Sérsic profile using GALFIT, a 2D parametric light profile fitting software tool to fit galaxy profiles (C. Y. Peng et al. 2002, 2010). We detail our fitting methodology in G+24. In short, we make image cutouts centered around the target galaxy and mask out all bright galaxies except the target galaxy and its bright neighbors in the 2D image. We then model the Sérsic profiles of all unmasked galaxies simultaneously, and we perform this modeling in two runs.

However, we make the following changes to our methodology to optimize the fitting process for galaxies in dense environments, since we specifically target QGs in the cluster cores in this study. When we fit the light profiles for the second time, we keep the parameter space free while using the fit results from the first round as input parameters. To address the intracluster light (ICL), we also allow GALFIT to fit the gradients in the background sky. If GALFIT fails to fit a galaxy, we re-fit the galaxy by masking out all bright sources in the background.

### 3.2. Simulations and Uncertainty Estimation

We conduct a series of simulations to check the robustness of our fitting algorithm and estimate the systematic errors in morphological measurements of galaxies. G+24 details the

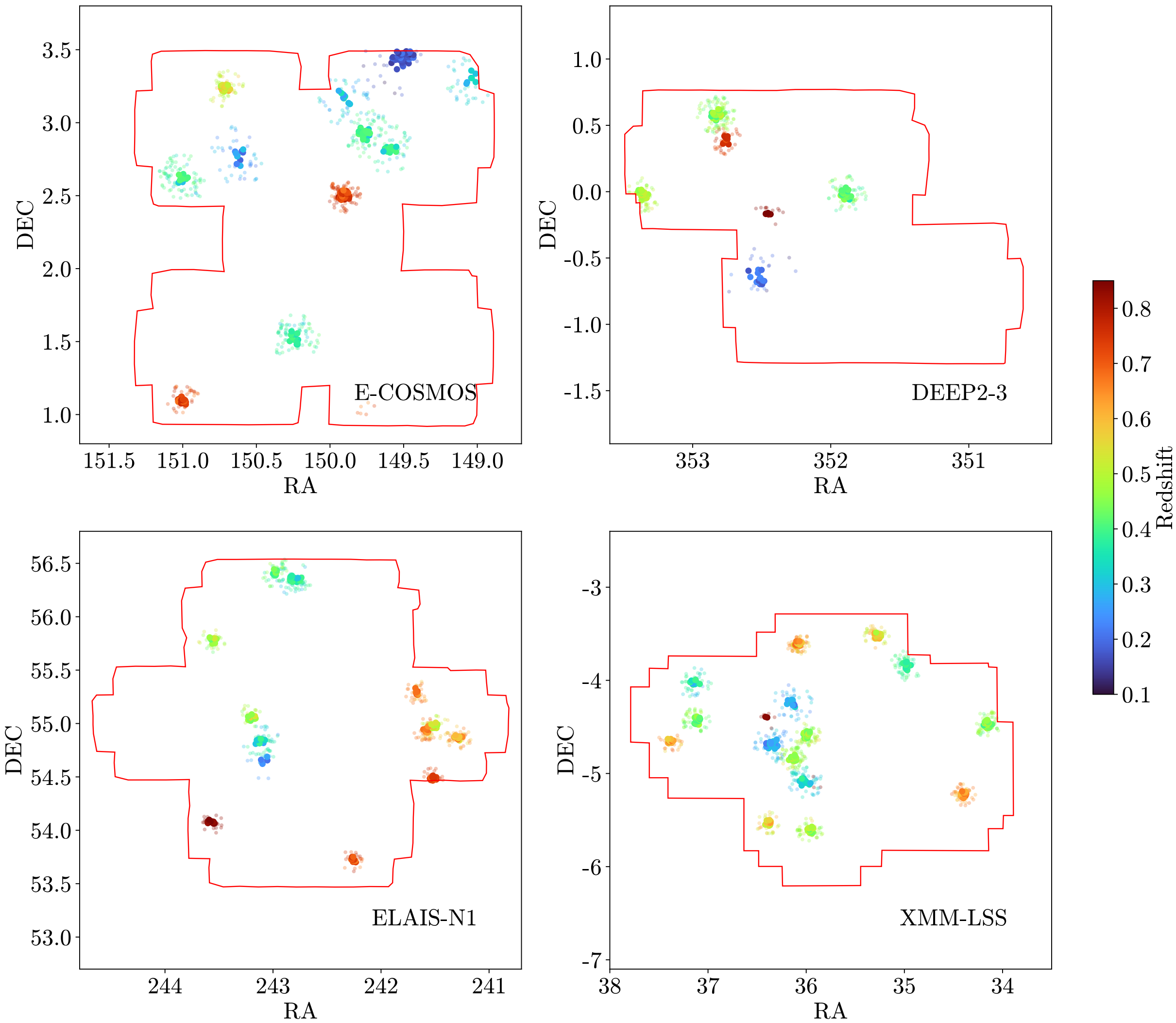

**Figure 2.** Selected QG members of galaxy clusters in four CLAUDS+HSC fields. Member galaxies are color-coded by their redshift. Large bright data points indicate the cluster QGs within a projected radius of 1 Mpc from the cluster center and the small faint points indicate the cluster QGs within 3 Mpc. The boundary of the CLAUDS survey in each field is shown in red.

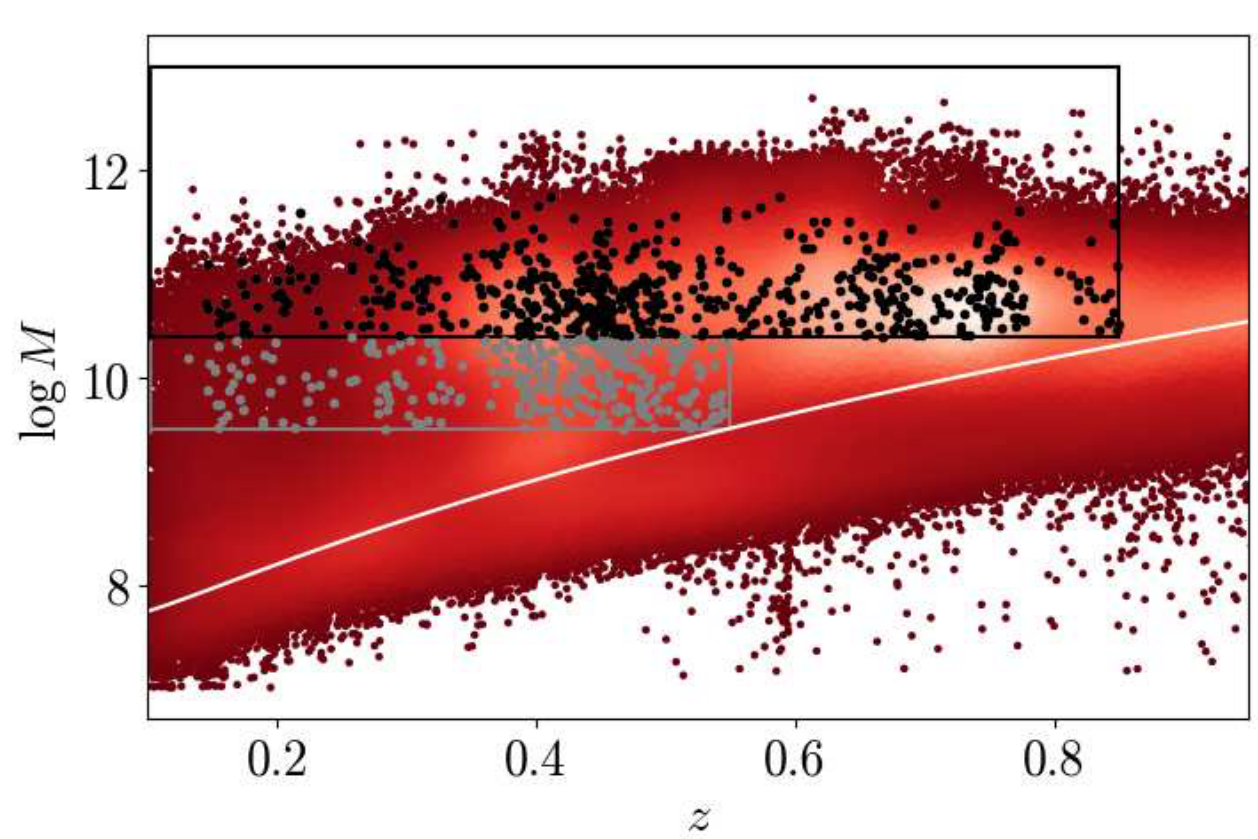

**Figure 3.** Stellar mass is plotted against redshift for QGs (red) in the parent catalog. The white curve shows the 90% completeness limit due to the magnitude cut we introduce at $i < 24$ for robust size measurements. The black box shows the QGs with mass, $\log M > 10.4$ above which our sample is mass-complete at $0.1 < z < 0.85$, whereas the gray box region shows the QGs with mass, $\log M > 9.5$ at $0.1 < z < 0.55$. The cluster QGs (see Section 2.3 for their selection) above the pivot mass are shown in black and below the pivot mass in gray.

**Table 1**
Sample Selection from the CLAUDS+HSC Data

| Category | Number |
|---|---|
| All objects ($i < 26$) | 7,493,735 |
| Galaxies with reliable photometry | 6,210,218 |
| $\log M > 9.5$ | 2,637,806 |
| $0.1 < z < 0.85$ | 297,213 |
| $i < 24$ | 289,165 |
| QGs | 87,815 |
| Field QGs | 86,658 (54,152) |
| Cluster QGs | 1,035 (575) |
| Reliable fits in UV | |
| Field QGs | 66,910 (46,034) |
| Cluster QGs | 798 (503) |
| Reliable fits in Optical | |
| Field QGs | 75,376 (47,336) |
| Cluster QGs | 909 (511) |
| Number of clusters | 47 |

**Note.** Numbers corresponding to massive ($\log M > 10.4$) QGs are provided inside brackets.

procedure for these simulations. In short, we generate and plant ∼50,000 point-spread function (PSF)-convolved mock images of galaxies into real data. To ensure that these artificial galaxies represent real galaxies, we model our galaxies using the distribution of morphological properties from the Zurich Structure and Morphology Catalog (M. T. Sargent et al. 2007; C. Scarlata et al. 2007). To ensure the diverse environments, we place these galaxies randomly in the data images. We then fit the Sérsic profiles of these galaxies using our methodology and use the fitting results to analyze how well we can recover their structural parameters based on their magnitude, size, Sérsic index, and axis ratio.

As in G+24, we use the results to quantify the dependence of systematic uncertainties in our measurements on the intrinsic properties of fitted galaxies. In addition, here, we also investigate the effect of the environment on the robustness and accuracy of our methodology. We investigate the aspect of environments by analyzing the simulation results as a function of local surface brightness (LSB). We estimate the LSB as

$$\mathrm{LSB} = -2.5 \times \log\left(\frac{\sum_x F_x}{900\pi}\right) \ [\mathrm{mag/arcsec}^2], \qquad (2)$$

where $\sum_x F_x$ denotes the sum of fluxes of all bright sources ($i < 28$) from the photometric catalog (G. Desprez et al. 2023) within a radius of 0′.5 centered around the simulated galaxy in the real data.

Figure 4 shows the success rate—the percentage of simulated galaxies fitted successfully—as a function of LSB and $i$-band magnitude of the simulated galaxy. The success rate is above 75% throughout the parameter space, except for very faint galaxies ($i > 24$). Motivated by this value, in what follows, we limit our analysis to galaxies brighter than $i = 24$. We also introduce similar cuts in other bands as well ($u < 25.5$, $g < 25$, $r < 24.5$, and $z < 24$).

The upper panel of Figure 4 illustrates the distribution of LSB values of the regions where we plant simulated mock galaxies. It shows that the environments of simulated galaxies cover a broad range of LSB values and that there is a long tail of probed LSB values that correspond to the densest cluster cores.

The success rate also drops in extremely bright areas of the sky ($\mathrm{LSB} < 25$ mag arcsec$^{-2}$), which, in our sample, mostly fall either on top or in the close proximity of BCGs. Since such bright regions are rare in our data, the number of mock galaxies planted in such areas is ∼1% (top panel of Figure 4). However, it is difficult to model light profiles of faint galaxies in such regions where the light from the BCG overwhelms, and the fitting results in such regions are unreliable. Therefore, we are confident that we can investigate the size evolution of galaxies irrespective of whether they reside in a dense or sparse environment, except maybe in the close proximity of BCGs.

To estimate total systematic errors in our fitting procedure, we follow the method described in G+24 and compare the input parameters of the artificial galaxies with those obtained from GALFIT's fits. To quantify the comparison, we define the relative difference for a given parameter $x$ as

$$\mathfrak{R}_d(x) = \frac{x\ (\mathrm{input}) - x\ (\mathrm{output})}{x\ (\mathrm{output})} \qquad (3)$$

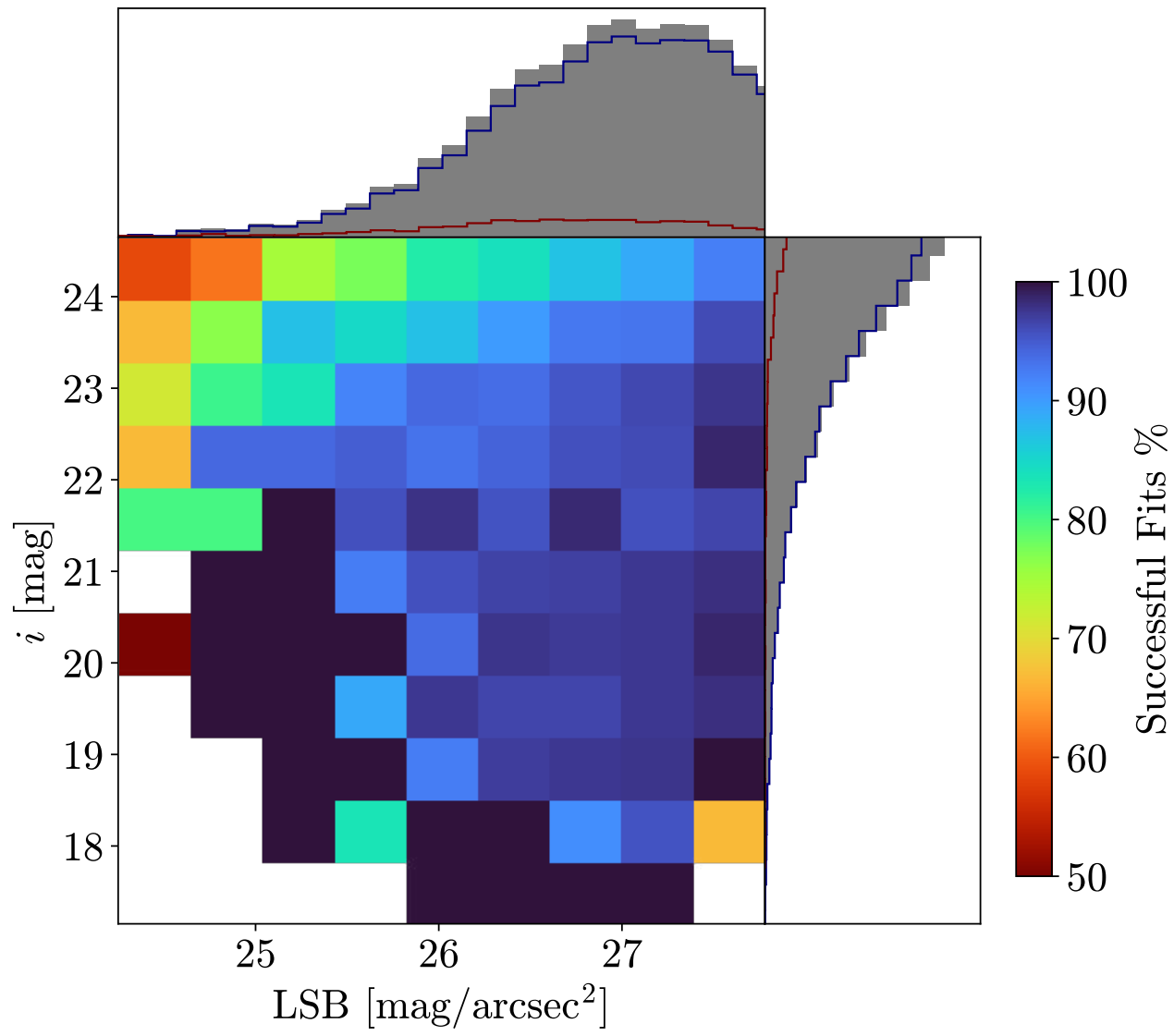

**Figure 4.** Heatmap showing the percentage of simulated galaxies we fit successfully using our algorithm in bins of $i$-band magnitude and local surface brightness (LSB). The success rate is mostly 75% or above throughout except at $i > 24$ and in regions of very high background with $\mathrm{LSB} < 25$ mag arcsec$^{-2}$. The white bins do not have any mock galaxies in the simulation. The 1D histogram on top shows the distribution of LSB of all simulated galaxies (gray), successful fits (blue), and failed fits (red). The histogram on the right shows similar distributions but for the $i$-band magnitude.

and investigate how this $\mathfrak{R}_d(x)$ varies with LSB, input magnitudes, and output structural parameters, $R_e$ and $n$. We then add the scatter in $\mathfrak{R}_d(x)$ in the 4D bins of LSB, magnitude, $R_e$, and $n$ as the systematic uncertainty to the GALFIT-generated random uncertainties in quadrature to compute the total error budget for our measurements. Thus, the uncertainties in Sérsic parameters adopted in this study include both random errors from GALFIT and systematic uncertainties estimated through simulations.

We show the distribution of $\mathfrak{R}_d(R_e)$ values for the $i$ band in Appendix A.[11] In general, the comparison of input and recovered structural parameters shows that we can reliably model galaxy light profiles using our pipeline and can approximate the distributions of $\mathfrak{R}_d(R_e)$ to Gaussian functions centered around 0. However, when we plant a very extended galaxy ($R_e > 1$") with a high Sérsic index ($n > 3$) in regions with LSB$< 26$, our algorithm yields a larger (>50%) size (Figure 9). These extended mock galaxies are planted in very crowded regions mostly over the images of BCGs. As such extended objects near BCGs are expected to be relatively rare compared to compact galaxies (e.g., S. E. Strom & K. M. Strom 1978; B. Moore et al. 1996, 1998; J. P. Madrid et al. 2010), we do not expect our pipeline's performance on such objects to affect this study.

### 3.3. Estimation of Morphological Parameters in Two Rest-frame Wavelengths

The rest-frame UV light is a proxy for star-forming regions within galaxies, while the rest-frame optical light traces the distribution of the bulk of the galaxy stellar population. Hence, using the fit results from the $U$ + griz bands, we estimate the

[11] The simulations in other bands also show similar results.

**Table 2**
Number of Galaxies with Successful Estimation of Sérsic Parameters in Two Rest Frames

| Redshift | 0.1–0.25 | 0.25–0.4 | 0.4–0.55 | 0.55–0.7 | 0.7–0.85 |
|---|---|---|---|---|---|
| Clusters | 6 | 20 | 25 | 21 | 11 |
| Rest-UV | | | | | |
| Field QG[a] | 3050 | 15785 | 16958 | 14712 | 16405 |
| Cluster QG | 70 | 181 | 307 | 135 | 105 |
| Cluster Outskirt QG[b] | 71 | 535 | 553 | 211 | 171 |
| Rest-Optical | | | | | |
| Field QG | 3374 | 15830 | 18855 | 17592 | 19725 |
| Cluster QG | 70 | 177 | 341 | 184 | 137 |
| Cluster Outskirt QG | 75 | 542 | 590 | 269 | 193 |

**Notes.** The cluster QGs are those within 1 Mpc projected distance from the cluster centers (BCGs). We classify those within 1–3 Mpc as cluster outskirt QGs. Some of the clusters have members in more than one redshift bin due to the photo-$z$ range we allow while selecting cluster members (see Section 2.3).
[a] Includes all QGs outside the inner 1 Mpc regions of the clusters.
[b] These are included in the field QG sample.

Sérsic structural parameters including $R_e$ and $n$ in two rest-frame wavelengths: 3000 Å (UV) and 5000 Å (optical). As detailed in G+24, for every galaxy, we take a weighted average of its structural parameters in the two nearest observed bands that either bracket or are close to the rest-frame 3000 and 5000 Å. Thus, the observed bands we use to estimate sizes in the same rest-frame regimes depend on the redshift of individual galaxies.

Out of 87,815 QGs, we successfully fitted UV profiles for 67,708 QGs and optical profiles for 76,285 QGs. QGs are generally fainter in the rest-frame UV than in the rest-frame visible regime, which drives the lower number of galaxies with rest-frame UV profiles. Even though we shift our magnitude selection cuts to fainter magnitudes in blue filters based on simulations we perform ($u < 25.5$, $g < 25$, and $r < 24.5$; see Section 3.2), we still miss around 10% of galaxies (both in clusters and field) because they are faint in the UV. However, this missing fraction is <3% at $\log M > 10.4$ (above the SMR pivot point). Table 2 details the number of galaxies with successful estimations of Sérsic models in rest frames 3000 and 5000 Å.

## 4. Results

Galaxies exhibit a well-established power-law relationship between stellar mass and effective radius, known as the size–mass relation (SMR) at least since $z \sim 5.5$ (e.g., S. Shen et al. 2003; I. Trujillo et al. 2004; R. J. Williams et al. 2010; A. van der Wel et al. 2014; N. Roy et al. 2018; L. A. Mowla et al. 2019; L. Kawinwanichakij et al. 2021; S. E. Cutler et al. 2022; A. George et al. 2024; K. Ito et al. 2024; M. Martorano et al. 2024; E. Ward et al. 2024). The SMR for both SFGs and QGs follows the form

$$\log R_e = \log R_0 + \alpha \log\left(\frac{M}{M_0}\right) + \sigma_{\log R_e}, \quad (4)$$

where $\alpha$ represents the slope, $R_0$ is the characteristic size at a fiducial mass $M_0$, and $\sigma_{\log R_e}$ is the intrinsic scatter in the linear regression. Following G+24, we adopt $M_0 = 5 \times 10^{10}\ M_\odot$. Notably, the SMR slope changes at a pivot point ($\log M \sim 10.3$) for QGs, while SFGs show a single slope across masses (e.g., A. van der Wel et al. 2014; L. Kawinwanichakij et al. 2021; A. George et al. 2024).

Appendix B describes how we determine the pivot point for QG SMR by fitting a smoothly broken double power law (Equation (B1)). The pivot points across the redshift bins ($z$: 0.1–0.25, 0.25–0.4, 0.4–0.55, 0.55–0.7, 0.7–0.85) range between $\log M = 10 - 10.4$ in both UV and optical wavelengths. Though we note that the sample used for the pivot point estimations is mass-complete in the first three redshift bins, we have a sizeable QG population in the other two redshift bins at $\log M = 9.5-10.4$ to fit the double power law. In addition, we do not find any significant evolution of the pivot point masses across the redshift bins.

Due to the relatively limited cluster QG sample, especially in the lowest- and the highest-redshift bins, we are unable to constrain the pivot points in their SMR as well as the field sample. Hence, for consistency, we adopt a universal pivot point, $\log M_0 = 10.4$, for field and cluster QGs across all redshift bins and at both rest-frame wavelengths. In the redshift bins where we have enough cluster QGs to fit the double power law, we compare the two pivot points and find that the cluster sample has a slightly larger pivot point mass that is still within $1\sigma$ of the value for the field samples.

In this work, we limit our analysis to QGs that are more massive than the pivot mass ($\log M > 10.4$) in field and cluster environments. We fit a linear regression model (Equation (4)) to these massive QGs using a Bayesian-based software package LINMIX.[12] The algorithm relies on the Markov Chain Monte Carlo method to obtain the best-fitting parameters ($R_0$ and $\alpha$). LINMIX incorporates the uncertainties in both mass and size measurements while fitting the data. The total error budget for size measurements that we provide as input for LINMIX includes our estimates of systematic errors from simulations in addition to much smaller random errors (Section 3.2).

Unlike our earlier work (G+24), we do not apply weights derived from the stellar mass function while fitting the SMR. Although the stellar mass function of QGs at $0 < z < 1$ is well constrained (e.g., A. Leauthaud et al. 2012; A. Muzzin et al. 2013; T. Moutard et al. 2016), the current literature lacks such an analysis of the QG population in the cluster cores over the complete redshift range. If we introduce the weights based on the same stellar mass function for both field and cluster galaxies (for example, A. Muzzin et al. 2013), the results of this work do not change. However, as the general (field) stellar mass function does not necessarily represent the cluster population, we omit the weights from the SMR fitting, consistently in both the field and the cluster sample.

Figure 5 illustrates the SMRs of field (black) and cluster (red) QGs above the pivot point in five redshift bins, with the shaded bands representing the uncertainty in the regression, while we provide the best-fit SMR parameters in Table 3. We also show the fiducial mass ($M = 5 \times 10^{10}\ M_\odot$) that we use to estimate the characteristic sizes of galaxies as vertical gray lines in each subpanel. This fixed mass at $\log M = 10.7$ represents roughly where the sample has the maximum number

[12] https://linmix.readthedocs.io/en/latest/src/linmix.html

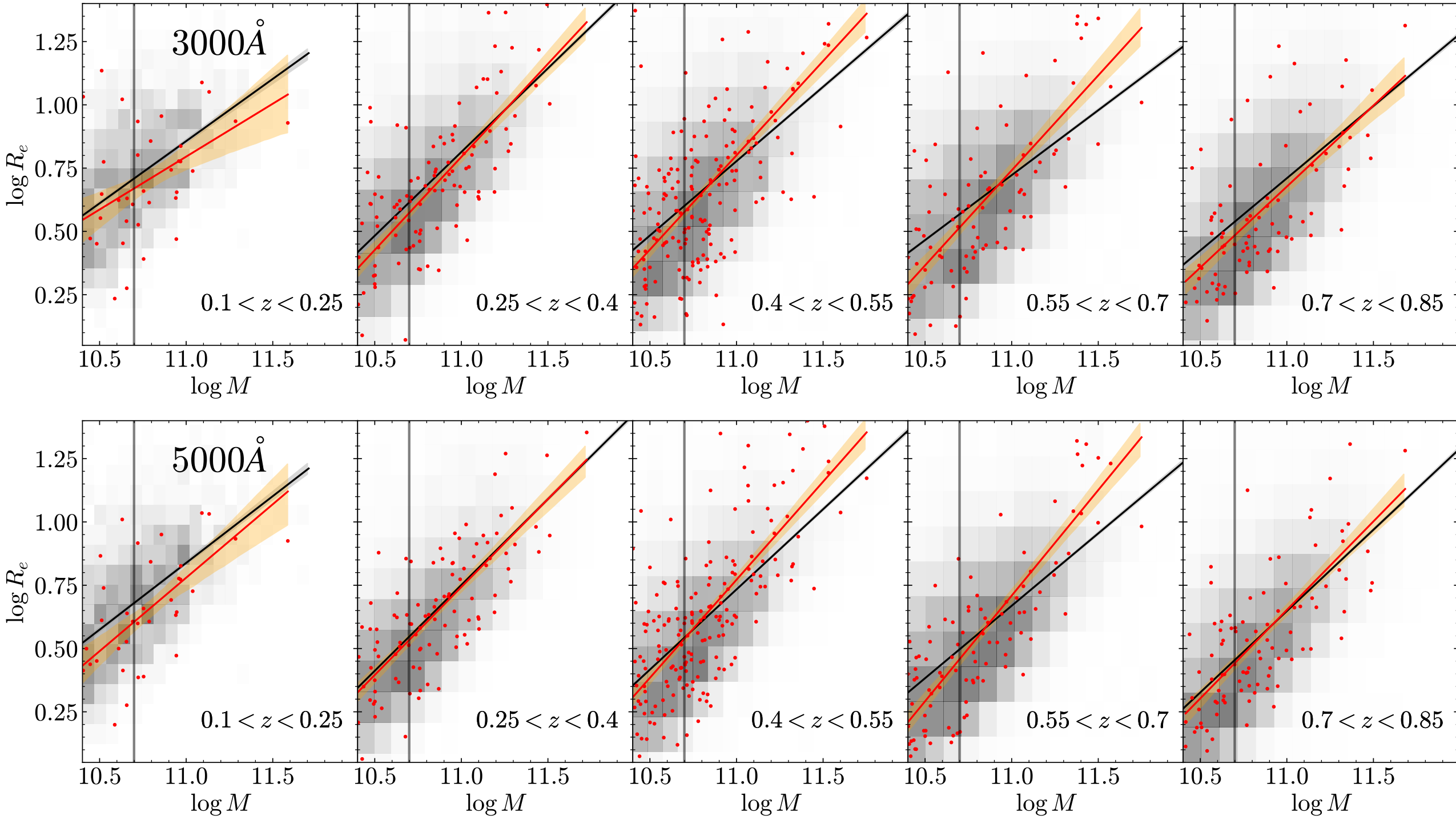

**Figure 5.** Size–mass relation (SMR) of massive QGs ($\log M > 10.4$) in clusters (red) and the field (black) in rest frame 3000 Å (top) and 5000 Å (bottom). We show the uncertainties in the SMR fitting in shaded bands. The red dots represent the cluster QGs used for fitting the SMR, and the gray histograms describe the distribution of field QGs. In each subpanel, the vertical gray line shows the fiducial mass, $M_0 = 5 \times 10^{10}\ M_\odot$, at which the characteristic size is estimated.

**Table 3**
The Best-fitting Parameters from the Single Power-law SMR Fits for Massive ($\log M > 10.4$) QGs in the Field and Cluster Environments in Two Rest-frame Wavelengths (Equation (4))

| | | | Rest Frame 3000 Å | | | |
|---|---|---|---|---|---|---|
| Environment | $z$ | median $z$ | $\alpha$ | $R_0$ (kpc) | $\sigma_{\log R_0}$ | $N$ |
| Field | 0.10–0.25 | 0.19 ± 0.00 | 0.49 ± 0.02 | 5.12 ± 0.06 | 0.17 ± 0.00 | 1361 |
| | 0.25–0.40 | 0.35 ± 0.00 | 0.66 ± 0.01 | 4.12 ± 0.02 | 0.20 ± 0.00 | 9342 |
| | 0.40–0.55 | 0.46 ± 0.00 | 0.59 ± 0.01 | 3.99 ± 0.02 | 0.22 ± 0.00 | 11027 |
| | 0.55–0.70 | 0.64 ± 0.00 | 0.51 ± 0.01 | 3.70 ± 0.03 | 0.23 ± 0.00 | 10405 |
| | 0.70–0.85 | 0.77 ± 0.00 | 0.57 ± 0.01 | 3.45 ± 0.02 | 0.21 ± 0.00 | 13899 |
| Cluster | 0.10–0.25 | 0.18 ± 0.01 | 0.45 ± 0.19 | 4.62 ± 0.47 | 0.21 ± 0.03 | 35 |
| | 0.25–0.40 | 0.35 ± 0.01 | 0.74 ± 0.07 | 3.73 ± 0.20 | 0.17 ± 0.01 | 105 |
| | 0.40–0.55 | 0.46 ± 0.00 | 0.74 ± 0.07 | 3.78 ± 0.16 | 0.21 ± 0.02 | 183 |
| | 0.55–0.70 | 0.62 ± 0.01 | 0.76 ± 0.08 | 3.26 ± 0.21 | 0.20 ± 0.02 | 92 |
| | 0.70–0.85 | 0.75 ± 0.00 | 0.63 ± 0.09 | 3.07 ± 0.23 | 0.18 ± 0.02 | 88 |
| | | | Rest Frame 5000 Å | | | |
| Environment | $z$ | median $z$ | $\alpha$ | $R_0$ (kpc) | $\sigma_{\log R_0}$ | N |
| Field | 0.10–0.25 | 0.19 ± 0.00 | 0.53 ± 0.02 | 4.78 ± 0.06 | 0.18 ± 0.00 | 1510 |
| | 0.25–0.40 | 0.35 ± 0.00 | 0.68 ± 0.01 | 3.52 ± 0.02 | 0.19 ± 0.00 | 9249 |
| | 0.40–0.55 | 0.46 ± 0.00 | 0.63 ± 0.01 | 3.49 ± 0.02 | 0.21 ± 0.00 | 11158 |
| | 0.55–0.70 | 0.64 ± 0.00 | 0.57 ± 0.01 | 3.14 ± 0.02 | 0.23 ± 0.00 | 10757 |
| | 0.70–0.85 | 0.77 ± 0.00 | 0.64 ± 0.01 | 2.85 ± 0.02 | 0.22 ± 0.00 | 14662 |
| Cluster | 0.1–0.25 | 0.18 ± 0.01 | 0.58 ± 0.16 | 3.98 ± 0.34 | 0.19 ± 0.03 | 35 |
| | 0.25–0.40 | 0.35 ± 0.01 | 0.69 ± 0.07 | 3.43 ± 0.16 | 0.15 ± 0.01 | 101 |
| | 0.40–0.55 | 0.46 ± 0.00 | 0.77 ± 0.06 | 3.45 ± 0.13 | 0.19 ± 0.01 | 180 |
| | 0.55–0.70 | 0.61 ± 0.01 | 0.84 ± 0.07 | 2.84 ± 0.14 | 0.18 ± 0.02 | 99 |
| | 0.70–0.85 | 0.75 ± 0.00 | 0.71 ± 0.08 | 2.72 ± 0.16 | 0.18 ± 0.02 | 96 |

**Note.** Uncertainties smaller than 0.005 are rounded as 0.00. The last column "$N$" provides the number of massive QGs used to fit the SMR.

of data points and is in line with similar works in the literature (e.g., A. van der Wel et al. 2014; L. A. Mowla et al. 2019; L. Kawinwanichakij et al. 2021; A. George et al. 2024).

In each redshift bin, $\log M \lesssim 11$ QGs are, on average, more compact in clusters than in the field (Figure 5). However, the trend reverses at very high masses ($\log M \gtrsim 11$), where we find that the average sizes of cluster QGs are larger than those of field QGs in both wavelengths. For example, at $z \sim 0.5$, the SMR size of QGs with mass $\log M = 11$ is 5% larger in clusters than in the field. Thus, the SMRs for QGs in clusters are steeper than those in the field. A notable exception to this trend is at the lowest-redshift bin, where the slopes are similar between the two environments, and the cluster QGs are consistently smaller than the field QGs. However, we cannot draw any strong conclusions here because we are limited by the small sample size (35 cluster QGs from five clusters) in this redshift bin (Table 3).

The characteristic sizes of cluster QGs (SMR zero-points) are in general smaller than their field counterparts across all redshifts (Figure 6, top panels). Even though we see this trend in both wavelengths, the difference is clearer in the rest-frame UV, where the characteristic sizes in all probed redshift bins are systematically smaller in the cluster environments. Moving from the lowest-redshift bin to the highest bin, the field QGs are larger than their cluster counterparts in rest-UV by $10.82\% \pm 11.38\%$, $10.46\% \pm 5.98\%$, $5.56\% \pm 4.52\%$, $13.50\% \pm 7.44\%$, and $12.38\% \pm 8.52\%$, respectively. Although differences are within the $1\sigma$–$2\sigma$ range in individual redshift bins, they are systematically in the same direction: cluster QGs at fiducial mass are smaller than the field QGs. We discuss the evolutionary trends of QG sizes in clusters versus the field in further detail in Section 5.

We also note that the characteristic sizes reported in this work differ from those in our pilot study in the central COSMOS field (G+24). These differences are in part due to the different redshift intervals probed by the two studies. This change in the intervals results in a lower median redshift of each bin in this work than in the pilot study and, therefore, larger characteristic sizes. However, major sources of discrepancies between the studies are (1) the differences in the photometric bands available for the SED fitting, which can affect the redshift and stellar mass estimation, (2) the selection of QGs using rest-frame colors, and (3) the photometry codes used (SExtractor versus hscPipe) that will propagate into mass measurement systematics. For the COSMOS field, we used the COSMOS2020 catalog (J. R. Weaver et al. 2023), with a much better wavelength coverage (X-ray to radio) than in the current work.

The difference described above does not affect the significance of the results presented here. The reason for this is that the current results are based on the comparison of two samples for which physical parameters are estimated from exactly the same set of available data. Our current analysis is based on relative comparison, and any systematic effects that we could not account for will have the same impact on both subsets.

We find that the cluster QGs exhibit steeper SMRs in both wavelengths than the field QGs, especially at $z > 0.4$. For example, at $0.55 < z < 0.7$, where the difference is maximum, the rest-UV SMR slopes in the field and cluster are $0.51 \pm 0.01$ and $0.76 \pm 0.08$, respectively. In the same redshift range, the slopes of rest-frame optical SMRs are $0.57 \pm 0.01$ and $0.84 \pm 0.07$, respectively, for QGs residing in the field and cluster environment.

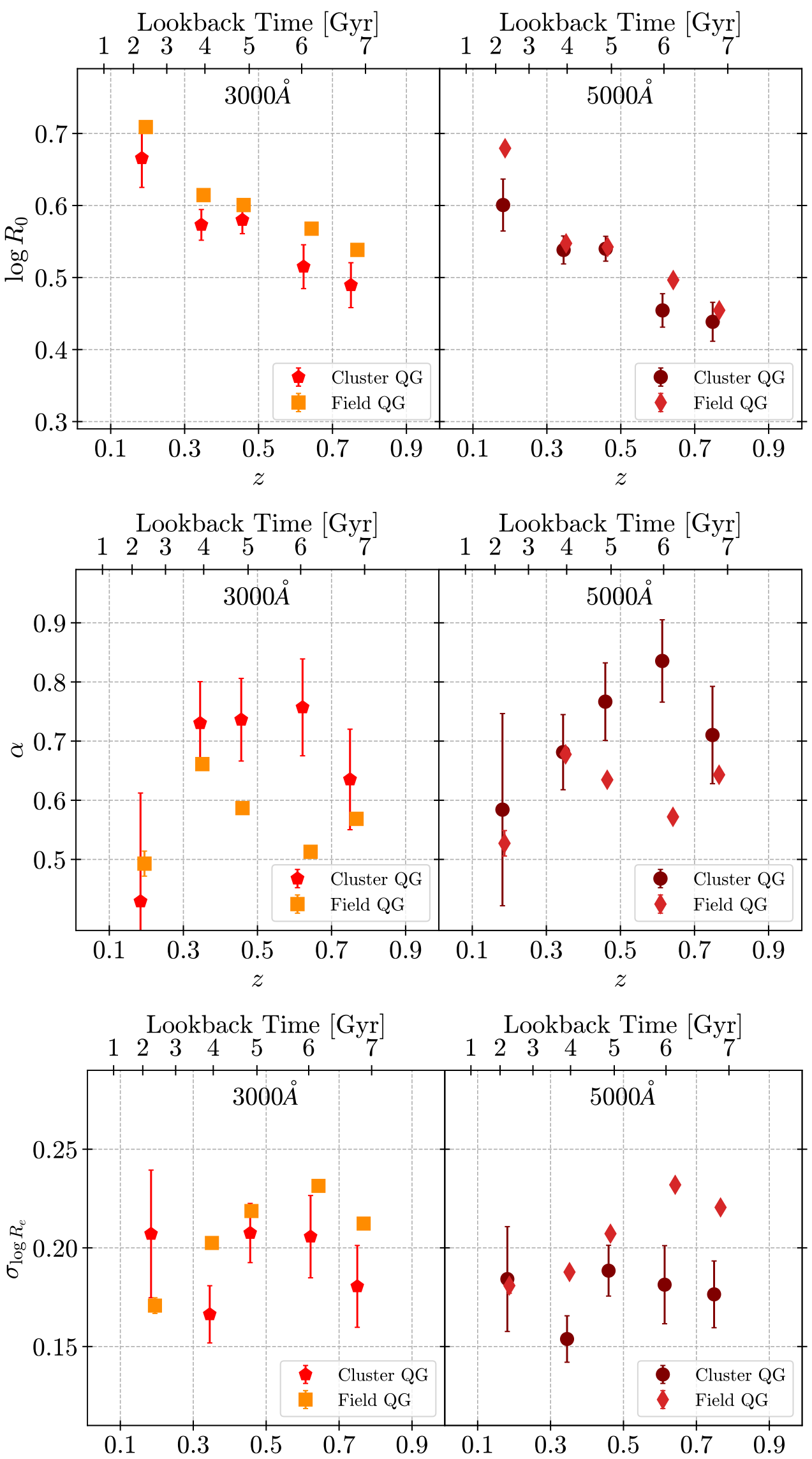

**Figure 6.** Top panel: the SMR intercepts (characteristic sizes at fiducial mass, $M = 5 \times 10^{10}\, M_\odot$) for cluster and field QGs in the rest-frame UV (left) and optical (right) wavelengths as a function of redshift. In both wavelengths, QG sizes decrease with redshift irrespective of environment. However, at any given redshift, the QGs are more extended in the field than in the clusters, and this difference is more prominent in the rest-frame UV than in the optical. Middle panel: the SMR slopes for cluster and field QGs measured in two rest-frame wavelengths and as a function of redshift. Although there is no strong correlation with $z$, the SMR in clusters tend to be steeper than in the field. Bottom panel: the intrinsic scatter in the SMR for cluster and field QGs. The details are the same as in the top panel. Cluster QG SMRs have generally lower scatter than their field counterparts.

However, we note that the slopes for the field QGs in this study are shallower than our previous findings in the pilot study in the COSMOS field (G+24), although the differences are statistically insignificant (within $1\sigma$–$2\sigma$). Possible reasons for this discrepancy in the SMR slopes are the differences in the redshift and stellar mass estimations between the two studies due to limitations in the available photometric bands for SED fitting, as discussed earlier. In fact, L. Kawinwanichakij et al. (2021) found even shallower slopes (0.34–0.44) in the HSC-SSP Wide area where they were limited to *grizy* bands for photo-$z$ and mass measurements.

The increase in the SMR slopes for cluster galaxies with respect to the general field in our sample deviates from a recent observational study at $z < 0.5$ using DESI Legacy Imaging Surveys by Z. Chen et al. (2024) that reports no difference in the SMR slope for QGs in dense environments. A major difference from our work is that they exclude BCGs from their cluster sample. Our cluster sample includes BCGs, which typically exhibit higher SMR slopes than other galaxies (Z. Chen et al. 2024; L. Yang et al. 2024). In fact, we find that the slopes of the cluster SMR become shallower (still steeper than the field SMR) when we remove the BCGs from our sample, although the major conclusions of this work remain unchanged. Nevertheless, we note that the discrepancy in slopes between Z. Chen et al. (2024) and this work is $\sim 1\sigma$. Furthermore, the differences in the data sets and methodology used for the size measurements between the works could also contribute to the observed differences in the SMR slopes.

Furthermore, we observe no significant redshift evolution in the SMR slopes for QGs in the field or clusters (middle panels in Figure 6), a result consistent with previous studies (e.g., R. F. J. van der Burg et al. 2016; P. Dimauro et al. 2019; L. A. Mowla et al. 2019; L. Kawinwanichakij et al. 2021; K. V. Nedkova et al. 2021; Z. Chen et al. 2024; A. George et al. 2024). The slopes for field QGs range between 0.5 and 0.7.

Although the intrinsic scatter in the SMR for field QGs increases with $z$, we do not observe a significant redshift evolution in the intrinsic scatter ($\sigma_{\log R_e}$) for cluster QGs across the two rest-frame wavelengths (bottom panels in Figure 6). Moreover, QGs in clusters consistently show lower intrinsic scatter in the relation between size and stellar mass than those in the field, except in the lowest-redshift bin.

To conclude, we find that the QG SMR differs between the field and clusters in rest-UV and optical, and this difference is observed across the redshift range probed in this study. Although we do not find a significant evolution in the slopes and intrinsic scatter in SMRs, the intercept (characteristic size of QGs with fiducial mass $5 \times 10^{10}\ M_{\odot}$) increases substantially with cosmic time for both field and cluster QGs. In the following section, we quantify and compare the evolution in the characteristic size of both quiescent subsets.

## 5. Discussion

We use the characteristic sizes of QGs with fiducial mass $5 \times 10^{10}\ M_{\odot}$ from the SMR to investigate the environmental impact on QG size evolution. We model this redshift evolution in characteristic sizes in the field and galaxy cluster cores in two rest-frame wavelengths (3000 and 5000 Å) using a power-law model, $R_e = R_e^0(1 + z)^{\beta}$, where $R_e^0$ is the characteristic size at $z = 0$. We provide the best-fit parameters in Table 4 and display the main results in Figure 7.

### 5.1. The Dependence of Size Evolution on Wavelength and Environment

The trends for field QGs align generally with those in the central COSMOS field study of G+24, suggesting consistency across the HSC Deep+UltraDeep fields. Our findings indicate that for a stellar mass of $5 \times 10^{10}\ M_{\odot}$, field QGs are more compact in optical than in UV light across the redshift range (the first panel in Figure 7).

**Table 4**

The Parameters of the Best-fit Power Law ($R_e = R_e^0(1 + z)^{\beta}$) to the Redshift Evolution of Characteristic QG Sizes in the Field and Clusters (Section 5)

| Environment | Rest-$\lambda$ | $R_e^0$ | $\beta$ |
|---|---|---|---|
| Field | 3000 Å | 5.82 ± 0.54 | −0.95 ± 0.23 |
| | 5000 Å | 5.61 ± 0.78 | −1.22 ± 0.33 |
| Cluster | 3000 Å | 5.38 ± 0.49 | −1.02 ± 0.22 |
| | 5000 Å | 4.77 ± 0.44 | −1.00 ± 0.23 |

The UV sizes of field QGs grow at a slower rate than their optical sizes ($\beta = -0.94 \pm 0.23$ versus $\beta = -1.22 \pm 0.33$). The average sizes of field QGs at $z = 0.85$ are 22% larger in UV than in optical but they become similar in both wavelengths by $z = 0$. In the rest-frame optical, the growth rate we observe here agrees with the growth rate ($\beta = -1.13 \pm 0.29$) we report in our pilot study (G+24). Although we find an even slower pace ($\beta = -0.69 \pm 0.28$) for the evolution of the UV sizes in the pilot work, the difference is within $1\sigma$. This small discrepancy can be attributed to a combination of the differences in the catalogs used (stellar mass estimates, SFG-QG separation, etc.) and the cosmic variance between the two studies (as the pilot study covers an area around 11 times smaller than this one).

While this wavelength dependence of QG sizes and their evolution was previously reported for the general QG population (G+24), our analysis is the first to reveal this trend separately in the field and cluster environments (the first and second panel of Figure 7, respectively). Notably, the size difference between measurements based on the rest-frame UV and optical bands is more pronounced for QGs in the field than for the ones in cluster cores. For example, at $z \sim 0.85$, a typical field QG of fiducial mass is 23% ($3\sigma$) larger in UV than in optical. This difference is only 12% (within $2\sigma$) in the cluster environment. Just like their field counterparts, cluster QGs are systematically smaller in the optical light than in the UV at all redshifts.

QGs in clusters grow as fast as field QGs in both rest-frame wavelength regimes. This fast pace of QG size growth in clusters is similar in rest-UV ($\beta = -1.02 \pm 0.22$) and optical ($\beta = -1.00 \pm 0.23$). Thus, the cluster QGs with fiducial mass $\log M = 10.7$ grows in size by 87% (85%) in the rest-UV (optical) over 6 Gyr of cosmic time (since $z = 0.85$).

In the right two panels of Figure 7, we further explore the environmental impact on QG sizes in UV and optical light. We find a marginal (within $1\sigma$) but systematic difference between field and cluster QGs at rest 5000 Å where the QGs in the clusters are smaller than their field counterparts (the fourth panel). For example, field QGs are $\sim$3% (15%) larger than cluster QGs in rest-optical at $z = 0.85$ ($z = 0.1$). This marginal size difference of cluster QGs at longer wavelengths qualitatively agrees with the results of several other studies in the literature (e.g., T. Valentinuzzi et al. 2010b; B. M. Poggianti et al. 2012; J. Matharu et al. 2019).

The key new result of our work is the difference we observe at the rest-frame 3000 Å. For the first time, we show that the cluster QGs are systematically smaller than field QGs at rest-3000 Å at all redshifts that we probe. This difference in UV sizes between the field and cluster QGs is larger and more prominent ($2\sigma$) than in the rest-optical (the third panel in

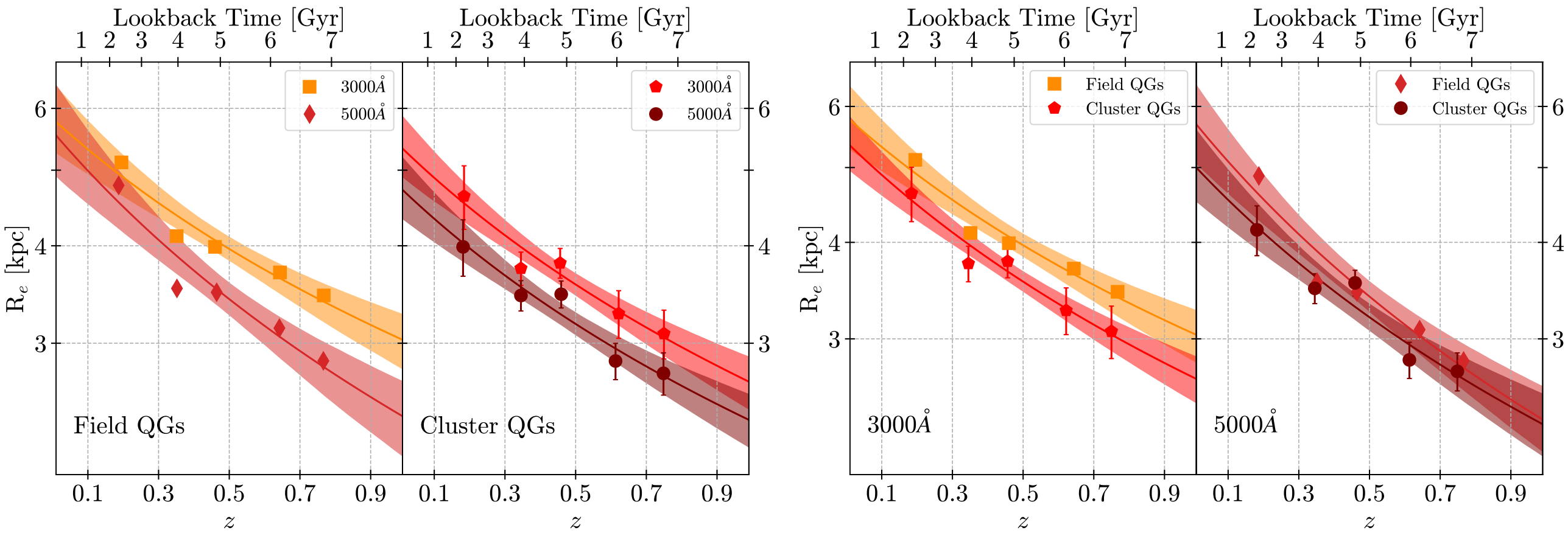

**Figure 7.** Redshift evolution of the characteristic sizes of cluster and field QGs in rest-frame UV and optical wavelengths. $R_e$ are characteristic sizes of $\log M = 10.7$ galaxies and are taken from the SMR fits. The best-fitting power-law functions are plotted as solid curves, with their uncertainties shown as shaded regions. Left panel: the first panel compares the size evolution of field QGs in UV (orange squares) and optical (maroon diamonds). The second panel compares the same for cluster QGs in UV (red pentagons) and optical (brown circles). Right panel: the first panel depicts the size evolution of field and cluster QGs in the rest-frame UV, and the second panel illustrates the same in the rest-frame optical. The rest of the details are the same as in the left panels.

Figure 7). At rest-3000 Å, the field QGs are ∼13% (9%) larger than the cluster QGs at $z = 0.85$ ($z = 0.1$).

These observed trends in size, especially in rest-UV, with environment may arise from several mechanisms. For QGs in the field, merger-driven growth plays a substantial role at $z < 1$. The rate of major mergers, which occur between galaxies of comparable mass (typically mass ratios >1:4), decreases substantially over cosmic time, rendering them insufficient for significant size evolution alone (C. J. Conselice et al. 2008; K. Bundy et al. 2009; C. López-Sanjuan et al. 2010; J. M. Lotz et al. 2011; S. Kaviraj et al. 2014), but minor mergers (mass ratios < 1:4) and gradual accretion remain more prominent, effectively expanding QG sizes in the field at later epochs (e.g., R. Bezanson et al. 2009; T. Naab et al. 2009; J. M. Lotz et al. 2011; A. van der Wel et al. 2014; K. A. Suess et al. 2020, 2023; D. J. Williams et al. 2024). Our earlier work confirms the critical role of minor mergers in driving QG size growth from $z \sim 1$ (G+24). A series of minor mergers increases the UV sizes of field QGs preferentially because low-mass companions (with mass ratio < 1:10) have bluer colors and lower mass-to-light ratios than their host galaxies (K. A. Suess et al. 2023).

In dense cluster environments, however, high velocity dispersions of galaxies (cluster dispersion) with respect to internal galaxy velocity dispersion inhibit merger-driven growth, as high-speed encounters reduce the likelihood of satellite galaxies merging efficiently (D. Merritt 1985; A. G. Delahaye et al. 2017). This lack of merger-driven growth is consistent with smaller sizes of cluster QGs compared to their field counterparts, especially in the rest frame 3000 Å (right panels in Figure 7). Minor mergers and accretion add stellar material to galaxy outskirts; these newly added stars have a range of properties in terms of mass, age and metallicity that reflect their origins. Generally, considering their origins, these accreted stars are observed to be bluer in color (K. A. Suess et al. 2023). Thus, minor mergers can cause QGs to appear more extended in the UV than in the optical because the in situ–formed stars of QGs are old and have a more centrally concentrated distribution than the ex situ population. Due to the lack of minor mergers among the satellite galaxies in the clusters, the UV and optical sizes of QGs are comparable in clusters but not in the field (left panels in Figure 7).

Although minor mergers may be uncommon, cluster galaxies are subject to violent galaxy–galaxy interactions (A. J. Allen & S. Yabushita 1984; B. Moore et al. 1996, 1998; C. Park & H. S. Hwang 2009; A. Saintonge et al. 2012) and tidal stripping (e.g., A. Toomre & J. Toomre 1972; D. Merritt 1983, 1985; G. Byrd & M. Valtonen 1990; J. I. Read et al. 2006; E. L. Łokas 2020). The tidally stripped stellar material forms ICL, whose stellar mass can amount up to ∼40% of the total stellar mass of a cluster (E. Giallongo et al. 2014; V. Presotto et al. 2014; M. Montes & I. Trujillo 2014, 2018; K. E. Furnell et al. 2021). In addition, most of the ICL originate from massive galaxies (T. DeMaio et al. 2018; M. Montes & I. Trujillo 2018; J. Matharu et al. 2019). For example, T. DeMaio et al. (2018) found that the 75% of the light emitted by the BCG and ICL inside clusters originates from disrupted massive cluster galaxies ($>10^{10.4}\ M_\odot$). Because tidal disruptions preferentially strip stars from galaxy outskirts, such events result in a reduction in galaxy sizes. This effect can explain why cluster QGs are smaller than their equally massive quiescent counterparts in the field (right panels in Figure 7). In general, galaxies are more extended in UV than in optical light (left panels in Figure 7). The impact of tidal stripping is, therefore, a larger difference in sizes between field and cluster QGs at shorter (UV) wavelengths (the two panels on the right in Figure 7).

Even though QGs in clusters are affected by environmental disruptions, we still observe a high rate of size growth ($\beta \sim 1$) for cluster QGs at both rest-frame wavelengths. As merger-driven growth is suppressed in cluster environments, the observed evolution requires additional driving mechanisms. A likely contributor is the addition of new QGs to the cluster populations. These added QGs can include both newly quenched galaxies within the clusters (newcomers/progenitor bias; R. P. Saglia et al. 2010; I. Damjanov et al. 2023; A. George et al. 2024) and the influx of field QGs into the galaxy clusters ("pre-processing"; Y. Fujita 2004; G. De Lucia et al. 2012; D. Olave-Rojas et al. 2018; F. Sarron et al. 2019; F. Piraino-Cerda et al. 2024). The progenitor bias may contribute to the fast size growth (especially in UV) observed in cluster interiors, as incoming new members provide an ongoing supply of younger QGs that appear larger in UV than older, more evolved cluster members. Similarly, the QGs accreted from the quiescent field population can have extended UV sizes akin to field QGs.

The contribution of these newly added QGs can potentially affect the size evolution of the cluster QG population. With time, galaxies in clusters lose their mass through tidal disruptions or merge with the BCGs in cluster centers (E. Hayashi et al. 2003; G. De Lucia & J. Blaizot 2007; M. Gu et al. 2018; L. Yang et al. 2024). Mass loss through tidal disruptions can result in the disappearance of some (smaller) cluster QGs from our mass-complete sample over time (S. Ghigna et al. 2000; A. V. Kravtsov et al. 2004). Through tidal forces, a galaxy of $\log M = 10$ can lose up to 2 orders of magnitude in its mass (A. V. Kravtsov et al. 2004). Thus, some of the cluster QGs with mass $\log M \sim 10.5$ at higher redshifts may disappear from our $\log M > 10.4$ sample at lower redshifts. This skews the population at the low-mass end of the observed cluster QG SMR toward more recently accreted QGs that are expected to have more extended light profiles than older cluster members (e.g., B. M. Poggianti et al. 2012). Along with the growth of cluster BCGs at the high-mass end, this continuous replacement of QGs at $\log M < 11$ could increase the SMR zero-point (i.e., characteristic size) with cosmic time.

Moreover, hierarchical structure formation models predict an increasing rate of accreted galaxies in clusters, because clusters grow via continuous accretion of galaxies from surrounding structures, leading to an increasingly dense cluster core over time (G. De Lucia et al. 2007; S. L. McGee et al. 2009; R. J. Smith et al. 2012). Simulations also predict that the majority of the subhalos in the massive clusters at $z = 0$ are accreted at $z < 1$ (L. Gao et al. 2004; F. Jiang & F. C. van den Bosch 2017; Y. M. Bahé et al. 2019). This cluster growth mechanism may foster rapid size evolution of the cluster galaxy population, especially at shorter wavelengths. First, newly accreted QGs tend to have extended UV sizes reminiscent of field QG properties. Second, the newly accreted SFGs continue to quench environmentally and are integrated into the existing cluster QG population (progenitor bias). We have shown that progenitor bias increases the average sizes of QGs in both wavelengths in our earlier work on field galaxies (G+24). Thus, the continued addition of these two types of QG subpopulations—field QGs and field SFGs that subsequently environmentally quench—to the cluster QG population may explain QG size evolution in clusters.

### 5.2. Potential Caveats

The limitations discussed in the pilot study (G+24), including age–metallicity degeneracy and the selection of quiescent galaxies, apply to this study as well. Extended UV sizes could reflect a more spatially distributed population of young and/or low-metallicity stars within these galaxies (B. Gustafsson 1989; D. Streich et al. 2014). The selection of SFGs and QGs depends on the methodology adopted (color–color diagrams, specific SFRs, morphology, etc.).

In addition, we rely on photometric data that could introduce contamination in our cluster QG sample ($r_c < 1$ Mpc). First, our sample can be affected by contamination from field QG interlopers due to uncertainties in the photo-$z$, even though we reduce its impact through red sequence selection (Section 2.3). In addition, our cluster QGs, although selected to lie within $r_c < 1$ Mpc, include a fraction of QGs from the outskirts ($r_c > 3$ Mpc) because the (photometric) redshift probability distribution of cluster members is typically broader than the size of clusters along the line of sight ($\pm 1000\ \mathrm{km\ s^{-1}}$; J. Sohn et al. 2021).

To understand the impact of contamination from field interlopers, we estimate the fraction of these interlopers in our cluster sample. The average projected number densities in the inner $r_c < 1$ Mpc, $1 < r_c < 2$ Mpc, and $2 < r_c < 3$ Mpc regions of our 47 clusters are 10.88, 1.07, and 0.765 $\log M > 10.4$ QGs per Mpc$^2$, respectively. If we assume all galaxies in the outer 2–3 Mpc region are field interlopers, it would give a limit of 7% contamination in our cluster QG sample (10% if we consider $1 < r_c < 2$ Mpc region). To gauge whether this contamination fraction has a significant effect, we replace 10% cluster QGs with field QGs that are matched in redshift and stellar mass. This additional, artificial contamination increases the characteristic sizes of the cluster QGs measured from the SMR diagram by 0.5%–1.3% in the rest-UV and 0.3%–0.6% in the rest-optical. This small increase does not affect any of the findings described in Section 5.1. Therefore, our results are not affected by interloper field QGs in our cluster sample.

Second, we investigate the effect of QGs from cluster outskirts in our sample due to the broad range of photometric redshift PDFs for potential cluster members. The discrepancy between the size of the cluster and the PDFs for cluster members may affect the physical sizes of cluster galaxies. To explore this effect, we place all cluster QGs for all clusters at the redshift of cluster center (spectroscopically confirmed) and re-fit size–mass relations. The effect on the characteristic sizes is negligible.

Therefore, we conclude that the selection of quiescent cluster members using the red sequence ensures purity of our selection (i.e., minimal contamination both from the field and from the mismatch in redshift). This is in line with M. D. Gladders & H. K. C. Yee (2000) who found that the contamination in the cluster sample selected using cluster red sequence is below 5%.

Finally, the results presented in this study are based on the characteristic sizes of QGs with fiducial mass $M = 5 \times 10^{10}\ M_\odot$. Our choice of the fiducial mass is based on the number distribution of QGs in the clusters and the field in all redshift bins and is in line with the literature including our pilot study (e.g., A. van der Wel et al. 2014; L. Kawinwanichakij et al. 2021; A. George et al. 2024). However, we recognize that cluster QGs with higher masses ($\log M \gtrsim 11$) have larger sizes than field QGs in some redshift bins (Figure 5) as also previously reported (e.g., L. Yang et al. 2024).

The cluster QG population at these high mass ranges ($\log M \gtrsim 11$) is dominated by BCGs, and their evolution is different from the rest of the cluster galaxies. BCGs grow in size by accreting material from satellite galaxies through mergers and tidal disruptions. The BCGs have steeper SMR than field and other cluster QGs (Z. Chen et al. 2024; L. Yang et al. 2024). In addition, Z. Chen et al. (2024) showed that the ultra-massive non-BCG QGs ($\log M > 11.2$) also have steeper SMR slopes than other cluster QGs. Steeper SMR slopes of ultra-massive QGs could indicate that these are not significantly affected by the processes discussed in Section 5.1, such as ram pressure stripping and tidal stripping. The presence of these ultra-massive and large galaxies steepens the overall

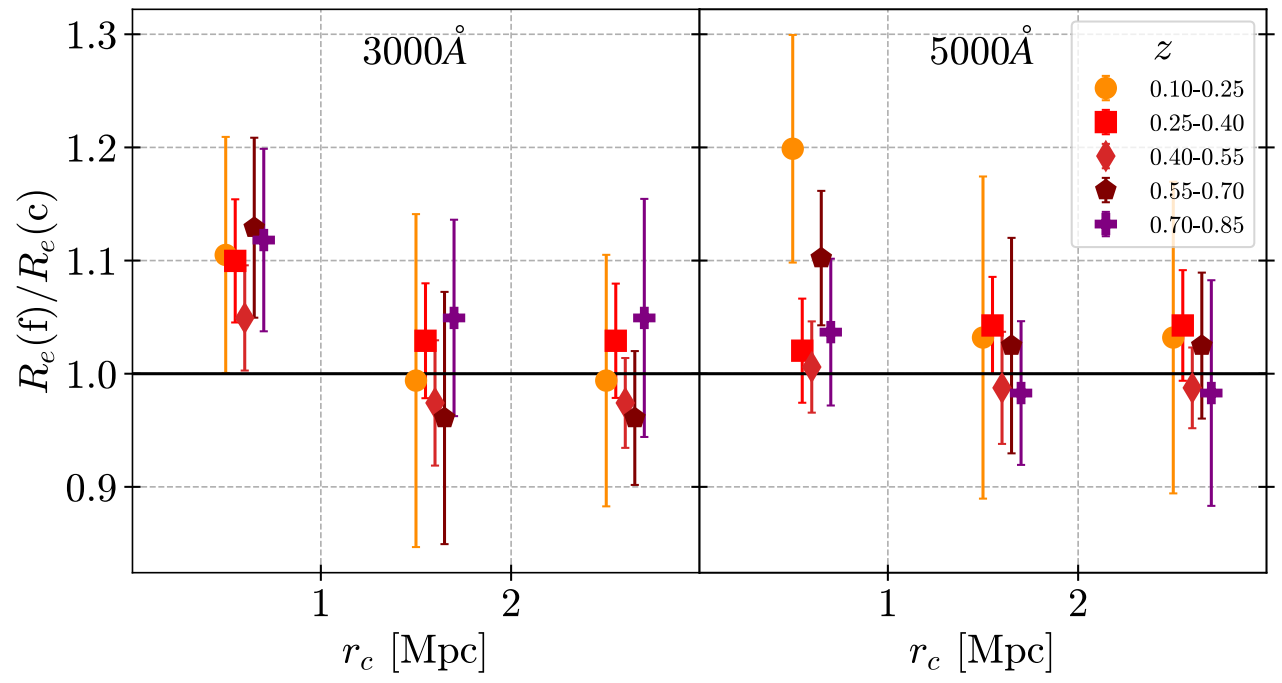

**Figure 8.** Field-to-cluster ratio of QG characteristic sizes as a function of the projected cluster-centric radius ($r_c$) in the rest-UV (left) and rest-optical (right). We show these ratios for cluster QGs at three different projected regions with their bootstrapped uncertainties. Except for QGs inside the central 1 Mpc region in the projected space, cluster and field QGs have similar sizes in both wavelengths.

SMR in the clusters such that the BCGs lie in the same SMR plane as other cluster QGs. Earlier works in the literature also show that the BCGs are part of the QG SMRs in the clusters, albeit making it steeper (M. Huertas-Company et al. 2013; Y. Yoon et al. 2017).

### *5.3. Comparison with Literature*

Earlier studies differ on the impact of dense environments in clusters on galaxy sizes. Cluster QGs are found to be comparable to the field galaxies in size (A. Rettura et al. 2010; A. B. Newman et al. 2014; T. Morishita et al. 2017; S. M. Sweet et al. 2017), but also smaller (T. Valentinuzzi et al. 2010a, 2010b; J. Matharu et al. 2019), or larger than their field counterparts (M. C. Cooper et al. 2012; C. Papovich et al. 2012; R. Bassett et al. 2013; C. Lani et al. 2013; L. Delaye et al. 2014; S. Andreon 2018; J. C. C. Chan et al. 2018; A. V. Afanasiev et al. 2023). Projection effects affect studies that analyze galaxy evolution in clusters across broad redshift ranges, as they face similar limitations with spectro-photometric data sets.

A comparison of our results with a plethora of previous studies listed above is not straightforward. This is because our work provides, for the first time, the results in two rest-frame wavelengths showing that the difference in size between field and cluster QGs depends on wavelength. The comparison with other studies is further complicated by the fact that they probe a range of cluster-centric regimes.

We find that the characteristic sizes of QGs in clusters increase with projected radial distance such that the sizes of QGs beyond 1 Mpc are similar to those of field QGs in both rest-frame wavelengths (Figure 8), although projection effects are more severe in the cluster outskirts. This pattern is consistent with known radial variations in galaxy properties, such as stellar age and morphology, driven by different evolutionary mechanisms that act at various distances from the cluster center (e.g., R. J. Smith et al. 2012; A. Biviano et al. 2013; A. Muzzin et al. 2014; D. S. Taranu et al. 2014; Y. L. Jaffé et al. 2015; I. Pintos-Castro et al. 2019; M. Pizzardo et al. 2023, 2024; M. Oxland et al. 2024). Therefore, this radial dependence in the QG field-to-cluster size ratio (Figure 8) may explain some of the discrepancies reported in the literature regarding the size comparison between field and cluster environments because the cluster-centric regimes probed by the studies vary significantly.

In addition, the choice of cluster sample selection methodology and probed ranges in stellar mass and redshift can alter the results. Our rigorous selection criteria based on cluster red sequence may preferentially remove post-starburst galaxies, as some of them are in the process of transitioning into the red sequence (M. M. Pawlik et al. 2018, 2019). Studying the SMR of post-starburst galaxies in clusters at $0.7 < z < 1.5$ using Hubble Space Telescope data, J. Matharu et al. (2020) showed that these galaxies lie on the large-size end of the QG SMR. Therefore, some of the discrepancies in the literature may arise from the differences in the fraction of the post-starburst galaxies in selected quiescent populations. Using ongoing and future large spectroscopic surveys, such as DESI (DESI Collaboration et al. 2024), PFS SSP (J. Greene et al. 2022), and WEAVE (S. Jin et al. 2024), we will be able to address this issue in detail with statistical samples of galaxy clusters with a large number of confirmed members.

Furthermore, the size difference between the field and cluster QGs varies with stellar mass because the slope of the SMR changes with the environment (Figures 5 and 6). Here, we adopt $\log M = 10.7$ QGs to compare their characteristic sizes in the clusters and the field. However, QGs have comparable sizes across environments at $10.7 \lesssim \log M \lesssim 11$ and the cluster QGs become larger than the field QGs at higher masses ($\log M \gtrsim 11$) in some redshift bins. As the stellar mass estimates depend on wavelength coverage and methodology, we need to be cautious when compiling and comparing the results of different studies.

Finally, previous works, in general, perform the size comparison between field and cluster galaxies in rest-optical. This is a key point, since this is where we find the differences between the field and cluster QGs to be small (within $1\sigma$ significance). The significant difference we find is in rest-UV—a wavelength regime that is not explored much in the literature—perhaps pointing to the fact that it is in the rest-UV where the processes that shape QG morphology have the most impact.

## 6. Conclusions

Using *Ugriz* images from 18.6 deg$^2$ of the CLAUDS+HSC survey, we study the morphological evolution of 87,815 massive ($\log M > 9.5$) quiescent galaxies (QGs) at $0.1 < z < 0.85$ in cluster and field environments. We successfully model the morphologies of 67,708 QGs in rest-frame UV and 76,285 QGs in rest-frame optical light using GALFIT. We study the impact of the environment on the morphological evolution of galaxies by classifying them into field and cluster QGs. From our sample, we identify 47 clusters and 1035 member QGs, selected within a cylindrical volume with a projected radius of 1 cMpc from the cluster center and depth defined by the photo-$z$ uncertainties, and refined by the cluster red sequence.

We fit the size–mass relation (SMR) of very massive ($\log M > 10.4$) field and cluster QGs in five redshift bins. Using the SMR characteristic sizes of QGs with fiducial mass, $M = 5 \times 10^{10} M_{\odot}$, we analyze the redshift evolution of QG sizes in the field and cluster at two rest-frame wavelengths. Our major findings are as follows:

1. The sizes of QGs are smaller in the rest-frame optical than in the UV in the field and in clusters. However, this

difference is more prominent for field QGs than cluster QGs.
2. Field QGs are significantly larger than cluster QGs in the rest-UV. In the rest-optical, this difference is marginal. In 3000 Å, the field QGs are ∼10% larger than the cluster QGs.
3. Sizes of cluster QGs grow at a fast rate with cosmic time, similar to those of field QGs ($\beta \sim 1$).

Unlike field QGs, which grow through minor mergers and accretion, cluster environments are not conducive to this merger-driven growth. In contrast, cluster-specific processes, such as tidal stripping, cause a decrease in QG sizes, especially in the rest-UV light. And yet, we observe a fast growth of the characteristic sizes of QGs in clusters on par with their field counterparts. In the absence of mergers, this fast growth is consistent with the cluster QG sizes growing through (1) the accretion of field QGs to the clusters that have larger sizes than cluster QGs and (2) the addition of recently quenched SFGs (that have larger sizes than QGs in general) to the cluster QG population (newcomers).

However, we note that, while the cluster QGs are smaller than field QGs at the fiducial mass explored in this work, this trend reverses at higher masses ($\log M > 11$) due to the dominance of the cluster BCGs. The growth mechanisms of BCGs differ from those of other cluster members and need separate analysis.

Thus, this study clearly shows that the sizes of QGs at mass $\log M = 10.7$ are smaller than their field counterparts in rest-UV. We also provide a plausible explanation for the evolution of QGs in cluster environments. With ongoing/future statistically large mass-limited spectroscopic surveys of clusters coming from instruments such as DESI (DESI Collaboration et al. 2024), Subaru Prime-Focus Spectrograph (PFS; J. Greene et al. 2022), VISTA 4MOST (R. S. de Jong et al. 2019), the Very Large Telescope MOONS (M. Cirasuolo et al. 2020), and WHT WEAVE (S. Jin et al. 2024), we can confirm the observations and explanations we put forward in this work using photometric data set. These large data sets will provide a considerably larger volume of cluster galaxies. For example, the DESI survey itself has identified 1.5 million clusters of galaxies (Z. L. Wen & J. L. Han 2024). With these large data sets, we can significantly reduce the uncertainties in the cluster SMR. The Subaru PFS, in particular, will target the same HSC Deep/UltraDeep fields. With the PFS, we will be able to confirm the cluster membership of our cluster sample and extend the analysis to cluster SFGs. This will enable probes of the growth of bulges in galaxies (A. George et al. 2024) and provide additional important constraints for the processes that drive the morphological evolution of galaxies in the densest environments.

## Acknowledgments

This research was supported by Discovery Grants to I.D. and M.S. from the Natural Sciences and Engineering Research Council (NSERC) of Canada. We utilize computational resources from ACENET, the Digital Research Alliance of Canada and CANFAR. We thank members of the extragalactic research group at Saint Mary's University, Canada, for their valuable insights and suggestions. We also thank the anonymous referee for providing thoughtful suggestions that helped us improve the clarity and completeness of the manuscript.

This work is based on data obtained and processed as part of the CFHT Large Area *U*-band Deep Survey (CLAUDS), which is a collaboration between astronomers from Canada, France, and China described in M. Sawicki et al. (2019). CLAUDS data products can be accessed from https://www.clauds.net. CLAUDS is based on observations obtained with MegaPrime/MegaCam, a joint project of CFHT and CEA/DAPNIA, at the CFHT which is operated by the National Research Council (NRC) of Canada, the Institut National des Science de l'Univers of the Centre National de la Recherche Scientifique (CNRS) of France, and the University of Hawaii. CLAUDS uses data obtained in part through the Telescope Access Program (TAP), which has been funded by the National Astronomical Observatories, the Chinese Academy of Sciences, and the Special Fund for Astronomy from the Ministry of Finance of China. CLAUDS uses data products from TERAPIX and the Canadian Astronomy Data Centre (CADC) and was carried out using resources from Compute Canada and the Canadian Advanced Network For Astrophysical Research (CANFAR).

This paper is also based on data collected at the Subaru Telescope and retrieved from the HSC data archive system, which is operated by the Subaru Telescope and Astronomy Data Center (ADC) at the National Astronomical Observatory of Japan. Data analysis was in part carried out with the cooperation of the Center for Computational Astrophysics (CfCA), National Astronomical Observatory of Japan. The Hyper Suprime-Cam (HSC) collaboration includes the astronomical communities of Japan and Taiwan, and Princeton University, USA. The Hyper Suprime-Cam (HSC) collaboration includes the astronomical communities of Japan and Taiwan, and Princeton University. The HSC instrumentation and software were developed by the National Astronomical Observatory of Japan (NAOJ), the Kavli Institute for the Physics and Mathematics of the Universe (Kavli IPMU), the University of Tokyo, the High Energy Accelerator Research Organization (KEK), the Academia Sinica Institute for Astronomy and Astrophysics in Taiwan (ASIAA), and Princeton University.

*Facilities:* CFHT (MegaCam), Subaru (HSC).

*Software:* NumPy (S. van der Walt et al. 2011; C. R. Harris et al. 2020), SciPy (P. Virtanen et al. 2020), Astropy (Astropy Collaboration et al. 2013, 2018, 2022), Photutils (L. Bradley et al. 2022), scikit-learn (F. Pedregosa et al. 2011), Matplotlib (J. D. Hunter 2007), GalfitPyWrap,[13] SExtractor (E. Bertin & S. Arnouts 1996), PSFEx (E. Bertin 2011), dynesty (J. S. Speagle 2020), Linmix,[14] GALFIT (C. Y. Peng et al. 2002, 2010).

## Appendix A
## Simulation Results

In Section 3.2, we describe the simulations we perform to estimate the systematic uncertainty present in the measurements of Sérsic parameters. We define a parameter, $\mathfrak{R}_d$ (relative difference; Equation (3)) to analyze the simulation results.

[13] https://github.com/Grillard/GalfitPyWrap

[14] https://linmix.readthedocs.io/en/latest/src/linmix.html

LSBD < 26

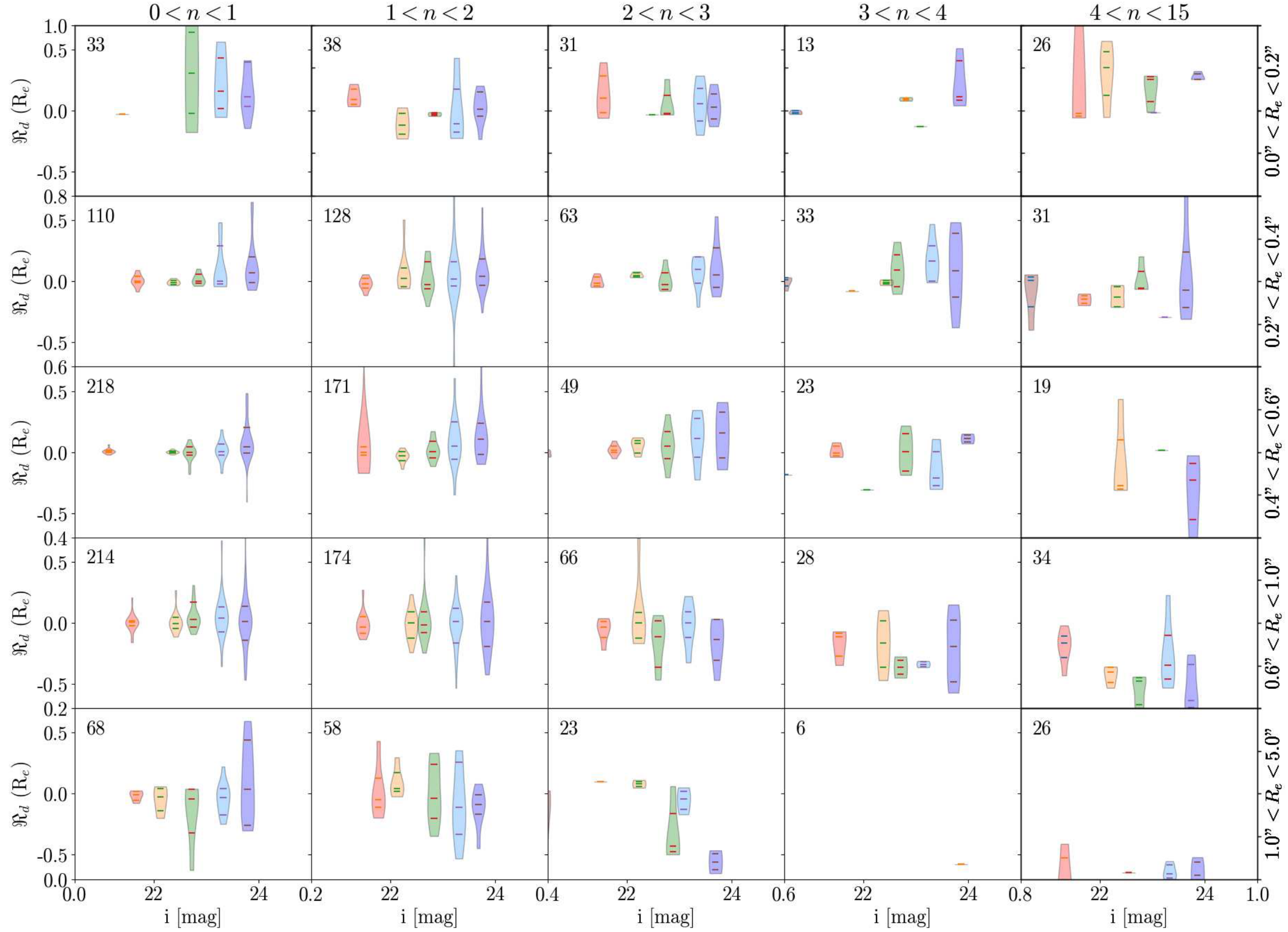

**Figure 9.** Distribution of relative difference in size measurements ($\mathfrak{R}_d(R_e)$; Equation (3)) in bins of *i*-band magnitudes, *Re* and *n*. All simulated galaxies are planted in the dense region, i.e., local surface brightness (LSB < 26).

We plot the distribution of $\mathfrak{R}_d(R_e)$ in 3D bins of *i*-band magnitudes, *Re* and *n* as violin plots in Figures 9–11. Figures 9, 10, and 11 show the distribution of relative difference for galaxies in three different local surface brightness (LSB) bins: LSB < 26, 26 < LSB < 27, and LSB < 27, respectively.

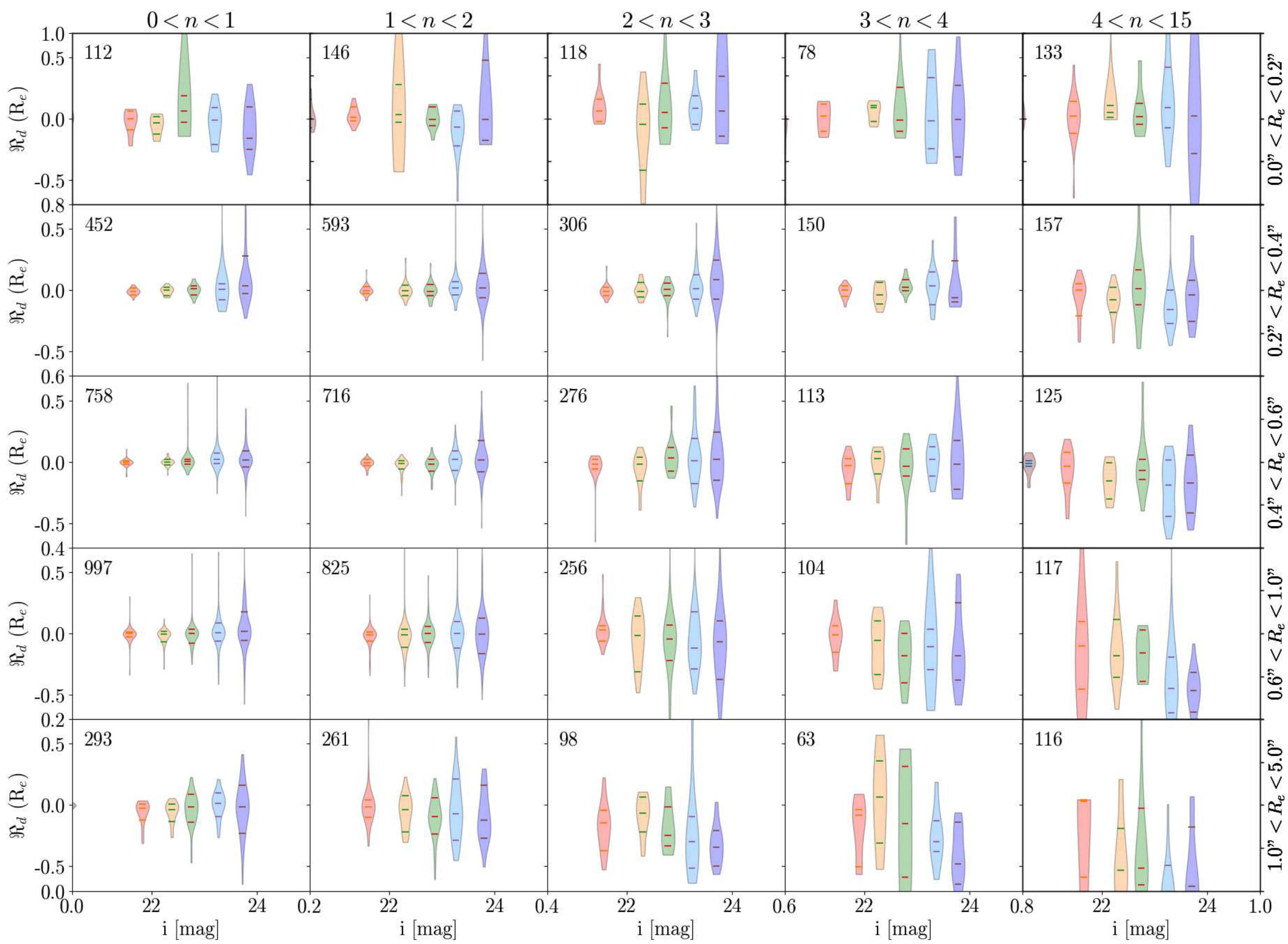

**Figure 10.** $\Re_d(R_e)$ for simulated galaxies with 26 < LSB < 27. Other details are the same as Figure 9.

LSBD > 27

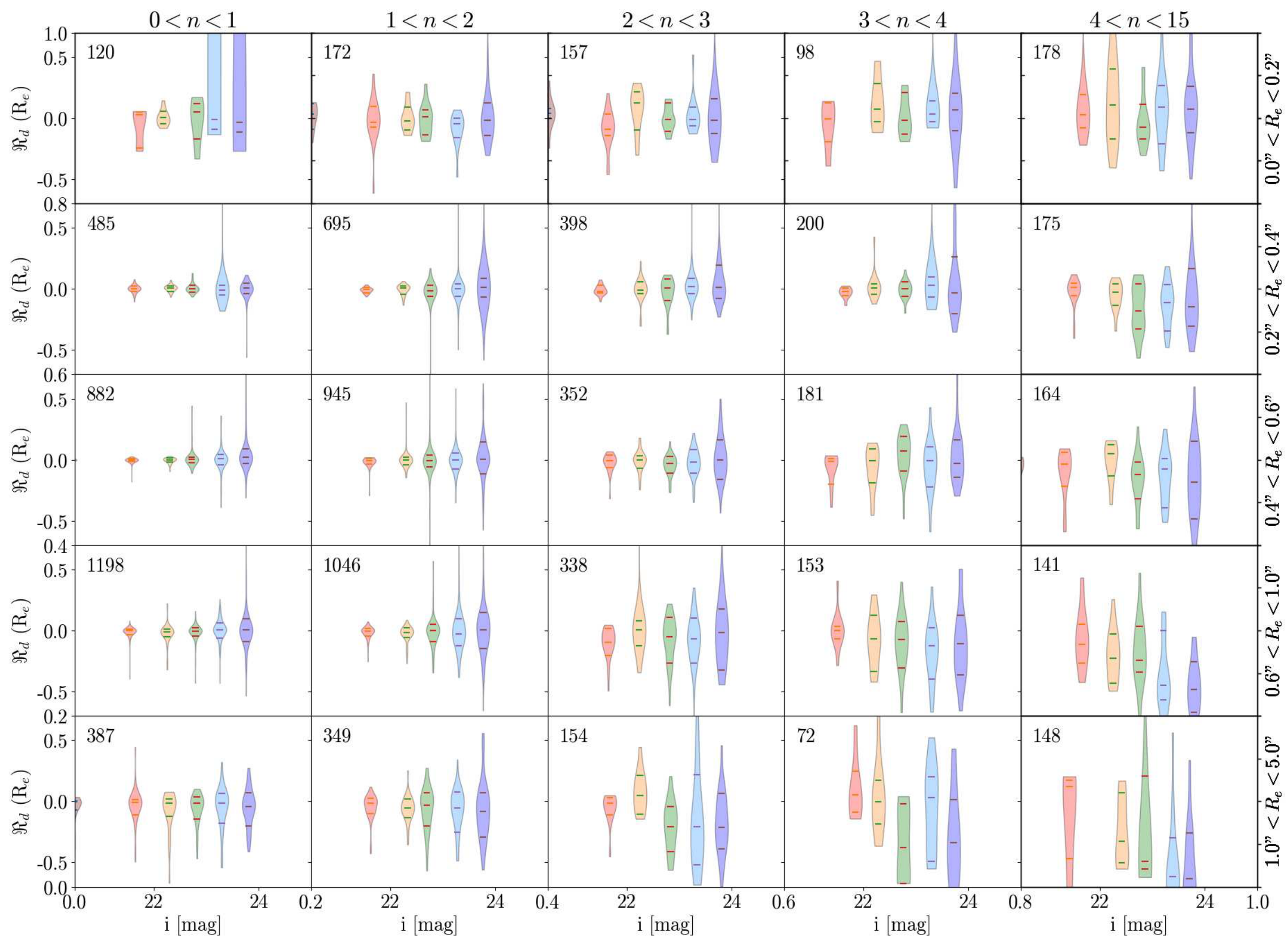

**Figure 11.** $\Re_d(R_e)$ for simulated galaxies with LSB > 27. Other details are the same as Figure 9.

## Appendix B Smoothly Broken Double Power Law

For the size–mass relation (SMR) of field QGs, as in G+24, we first fit a smoothly broken double power law,

$$R(M) = R_p\left(\frac{M}{M_p}\right)^{\alpha}\left[\frac{1}{2}\left(1+\frac{M}{M_p}\right)^{\delta}\right]^{\frac{\beta-\alpha}{\delta}}, \tag{B1}$$

where $M_p$ is the pivot stellar mass at which the slope changes, $R_p$ is the effective radius at the pivot stellar mass, $\alpha$ is the slope of the SMR at the low-mass end, $\beta$ is the slope at the high-mass end, and $\delta$ is the smoothing factor. We employ DYNESTY,[15] a PYTHON package designed to fit the data using Bayesian inference (J. S. Speagle 2020), following the methodology in G+24. To ensure that the contributions from different stellar mass ranges to the SMR are similar, we weigh the likelihood by a factor inversely proportional to the galaxy number density at a given stellar mass and redshift. We adopt this number density from A. Muzzin et al. (2013).

[15] https://zenodo.org/record/7832419

## ORCID iDs

Angelo George https://orcid.org/0009-0009-2742-7702
Ivana Damjanov https://orcid.org/0000-0003-4797-5246
Marcin Sawicki https://orcid.org/0000-0002-7712-7857
Guillaume Desprez https://orcid.org/0000-0001-8325-1742
Marianna Annunziatella https://orcid.org/0000-0002-8053-8040
Stéphane Arnouts https://orcid.org/0009-0008-8053-2512
Stephen Gwyn https://orcid.org/0000-0001-8221-8406
Danilo Marchesini https://orcid.org/0000-0001-9002-3502
Thibaud Moutard https://orcid.org/0000-0002-3305-9901
Anna Sajina https://orcid.org/0000-0002-1917-1200